\documentclass[fleqn, usenatbib, useAMS]{mnras}

\usepackage{newtxtext,newtxmath}
\usepackage[T1]{fontenc}
\usepackage{ae,aecompl}
\usepackage[utf8]{inputenc}

\usepackage{graphicx}	
\usepackage{amsmath}	
\usepackage{amssymb}	
\usepackage{bm}		    

\newcommand{\partderiv}[2]{\frac{\partial #1}{\partial #2}}
\newcommand{\Tr}{\mathrm{Tr}}
\newcommand{\lhf}{\mathcal{L}}
\defcitealias{palanque-delabrouille_one-dimensional_2013}{PD13}
\defcitealias{walther_new_2017}{W17}
\defcitealias{mcdonald_ly$upalpha$_2006}{M06}

\title[Optimal $P_{\mathrm{1D}}$ Estimate]{Optimal 1D Ly-$\alpha$ Forest Power Spectrum Estimation I: DESI-Lite Spectra}

\author[N. G. Kara\c cayl\i, A. Font-Ribera and N. Padmanabhan]{Naim G\" oksel Kara\c cayl\i$^{1}$, Andreu Font-Ribera$^{2, 3}$ and Nikhil Padmanabhan$^{1, 4}$
\\
$^{1}$Department of Physics, Yale University, New Haven, CT, USA\\
$^{2}$Department of Physics and Astronomy, University College London, London, UK\\
$^{3}$Institut de F\'isica d'Altes Energies (IFAE), The Barcelona Institute of Science and Technology, 08193 Bellaterra (Barcelona), Spain\\
$^{4}$Department of Astronomy, Yale University, New Haven, CT, USA}

\date{Accepted XXX. Received YYY; in original form ZZZ}

\pubyear{2020}

\begin{document}
\label{firstpage}
\pagerange{\pageref{firstpage}--\pageref{lastpage}}
\maketitle

\begin{abstract}
The 1D Ly-$\alpha$ forest flux power spectrum $P_{\mathrm{1D}}$ is sensitive to scales smaller than a typical galaxy survey, and hence ties to the intergalactic medium's thermal state, suppression from neutrino masses and new dark matter models. It has emerged as a competitive framework to study new physics, but also has come with various challenges and systematic errors in analysis. In this work, we revisit the optimal quadratic estimator for $P_{\mathrm{1D}}$, which is robust against the relevant problems such as pixel masking, time evolution within spectrum and quasar continuum errors. We further improve the estimator by introducing a fiducial power spectrum, which enables us to extract more information by alleviating the discreteness of band powers. We meticulously apply our method to synthetic DESI spectra and demonstrate how the estimator overcomes each challenge. We further apply an optimisation scheme that approximates the Fisher matrix to three elements per row and reduces computation time by 60\%. We show that we can achieve percent precision in $P_{\mathrm{1D}}$ with 5-year DESI data in the absence of systematics and provide forecasts for different spectral qualities.
\end{abstract}




\section{Introduction}
Through absorption lines in quasar spectra, the Ly-$\alpha$ forest technique can probe matter in vast volumes far into the past and at smaller scales than galaxy surveys, that are shaped by the thermal state of the gas and reionization history of the universe \citep{hui_equation_1997, gnedin_probing_1998}. Connecting flux fluctuations in quasar spectra to physical parameters relies on multiple demanding steps in a typical  analysis. The first step is to summarize the statistical information contained in millions of pixels across thousands of spectra using the correlation function or the power spectrum. Second, one relies on numerical simulations to relate the matter fluctuations to the neutral hydrogen which reionizes until $z\sim 6$ to obtain mock quasar spectra. The physical parameters are then mapped to the statistics using large numbers of these mocks with different parameters. The final step constrains physical parameters by performing a likelihood analysis on the observed statistics using this mapping and a prior.

The Ly-$\alpha$ forest technique already proved to be fruitful. The Extended Baryon Oscillation Spectroscopic Survey (eBOSS) \citep{dawson_sdss-iv_2016} and its predecessors successfully measured baryon acoustic oscillations in large-scale 3D correlations of the Ly-$\alpha$ forest \citep{slosar_lyman-alpha_2011, slosar_measurement_2013, busca_baryon_2013, font-ribera_quasar-lyman_2014, delubac_bao_2015,duMasdesBourboux_bao_2017, bautista_julian_e._measurement_2017, blomqvist_baryon_2019, deSainteAgathe_bao_2019}. It has also emerged as a promising tool to investigate intergalactic medium (IGM) thermal evolution \citep{boera_reionization_igm_2019, walther_new_2019}, to constrain neutrino masses \citep{croft_cosmological_1999, seljak_neutrino_2006, palanque-delabrouille_constraint_2015,palanque-delabrouille_neutrino_2015, yeche_constraints_2017} and to probe the nature of dark matter \citep{boyarsky_lyman-$upalpha$_2009, viel_warm_dm_2013, baur_lyman-alpha_2016, irsic_fuzzy_lya_2017, garzilli_lyman-_2019}. 

The line-of-sight flux power spectrum of the Ly-$\alpha$ forest has been at the frontier in new physics by being sensitive to medium to small scales. \citet{mcdonald_ly$upalpha$_2006}, \citet{palanque-delabrouille_one-dimensional_2013} and \citet{chabanier_one-dimensional_2019} measured this 1D power spectrum at large to medium scales ($0.001$ s km$^{-1} \leq k \leq 0.02$ s km$^{-1}$) using thousands of quasar spectra, whereas \citet{viel_warm_dm_2013}, \citet{irsic_lya_xq100_2017}, \citet{walther_new_2017}, \citet{boera_reionization_igm_2019} and \citet{day_power_2019} pushed the measurement to smaller scales ($k \leq 0.1$ s km$^{-1}$) using few but high-resolution spectra.

These recent works utilize three different methods in power spectrum estimation. \citeauthor{palanque-delabrouille_one-dimensional_2013} and \citeauthor{chabanier_one-dimensional_2019} applies fast Fourier transforms (FFT), which require all pixels to be present, equally spaced and have uniform noise and resolution. These conditions are rarely met in real spectra due to masking of sky emission lines, high column density absorbers (HCD), bad pixels and sometimes metal contamination. As a result, \citeauthor{chabanier_one-dimensional_2019} apply up to 20\% corrections to their power spectrum estimates. In their small-scale measurement, \citeauthor{walther_new_2017} and \citeauthor{day_power_2019} use Lomb-Scargle periodogram \citep{lomb_least-squares_1976, scargle_studies_1982} as this method allows for masking. However, neither method is able to weight pixels with respective noise estimates and cannot account for the time evolution within a spectrum without splitting the spectrum into multiple chunks; and both are limited by S/N and resolution in their quasar samples. Third method is the likelihood maximization. \citeauthor{palanque-delabrouille_one-dimensional_2013} also implemented a direct maximization of the likelihood function, which they found sensitive to the implementation details and noise in spectra. Our focus in this paper is the faster and more stable quadratic maximum likelihood estimator (QMLE), which \citeauthor{mcdonald_ly$upalpha$_2006} applied while splitting the spectra into two chunks because of computational limitations. However, with recent developments in computing, QMLE promises its full strength by careful application to 1D power spectrum.

Our work focuses on developing and applying an improved QMLE to measure the 1D power spectrum. We meticulously apply the formalism to maximize the information extraction. Our method readily handles the redshift evolution by interpolating pixel pairs to two redshift bins. This enables us to keep the full forest and all pixel pairs. We also introduce a baseline estimate that improves the accuracy by alleviating discrete band powers. As The Dark Energy Spectroscopic Instrument (DESI) \citep{levi_desi_2013, desi_collaboration_desi_2016} comes online to find abundance of quasar spectra, our improved QMLE can exploit the most information and make the best measurement.

This paper is organized under six sections. We summarize relevant general and specific formula for QMLE in section~\ref{sec:theory}. This section also briefly discusses a continuum limit to clarify what QMLE actually yields. Section~\ref{sec:implementation} outlines the algorithm and further provides details on our implementation, including validation with synthetic spectra. We then move to applying our method to simulated simple DESI data in section~\ref{sec:desilite_spectra}. In this section, we examine the effects of gaps and continuum errors, the advantage of fiducial power, Fisher matrix approximation and five-year forecasts for DESI with different spectral qualities. Section~\ref{sec:discussion} discusses the finer details of our method, such as interpretation of QMLE results, its advantages and possible problems. Finally, we summarize this work in section~\ref{sec:summary}.

\section{Method}
\label{sec:theory}
The optimal power spectrum estimator was extensively studied by \citet{hamilton_towards_1997},  \citet{tegmark_karhunen-loeve_1997}, \citet{tegmark_measuring_1998} and \citet{seljak_weak_1998}. We first assign a Gaussian probability distribution that depends on parameters $\theta_{\alpha}$ for measuring a data set $\bm{x}$. We define the likelihood function $\lhf$ as twice the logarithm of this probability and pursue the most likely parameters $\hat{\theta}_{\alpha}$ for a fixed $\bm{x}$.
\begin{equation}
    \lhf(\bm{x}; \theta_{\alpha}) = -\ln\det\mathbf{C} - \bm{x}^\mathrm{T}\mathbf{C}^{-1} \bm{x},
\end{equation}
where $\textbf{C} = \textbf{C}(\theta_{\alpha}) \equiv \langle \bm{x} \bm{x}^\mathrm{T} \rangle$. The most likely parameters $\bm{\hat\theta}$ can be found by maximizing this likelihood function: $\lhf_{,\alpha}(\bm{\hat\theta})=0$, where comma represents a partial derivative, which can be iteratively solved using the Newton-Raphson method:
\begin{equation}
    \hat\theta^{(X+1)}_{\alpha} = \hat\theta^{(X)}_{\alpha} - \sum_{\alpha'} \left.\left\langle\lhf_{,\alpha\alpha'} \right\rangle^{-1}\right|_{\bm{\hat\theta}^{(X)}} \lhf_{,\alpha'}(\bm{\hat\theta}^{(X)}),
\end{equation}
where $X$ is the iteration number. Note that we compute the ensemble average of the second derivative (which is the Fisher matrix) instead of using the full curvature matrix as has been the convention.

Our goal is to estimate the power spectrum of the observed spectra, so we take $\theta_{\alpha}$ to be the power spectrum estimates. Furthermore, we would like to estimate deviations from a fiducial power spectrum such that $P(k, z) = P_{\mathrm{fid}}(k, z) + \sum_{\alpha} w_{\alpha}(k, z) \theta_{\alpha}$, where $w_{\alpha}(k, z)$ are the functional forms for deviations and $\theta_{\alpha}$ are the amplitudes \citep{font-ribera_how_2018}. Then, the covariance matrix is the sum of signal and noise as usual, $\mathbf C = \mathbf{S}+\mathbf{N}$; and by extension $\mathbf{S} = \mathbf{S}_{\mathrm{fid}} + \sum_{\alpha} \mathbf{Q}_{\alpha} \theta_{\alpha}$, where $\mathbf{Q}_{\alpha} = \partial \mathbf{C} / \partial \theta_{\alpha}$. The fiducial power needs to be subtracted from the estimate. We denote its contribution  by $t_{\alpha}$ below, which can be immediately calculated by substituting $\mathbf{N} \rightarrow \mathbf{N} + \mathbf{S}_{\mathrm{fid}}$ to the quadratic estimator equation in the references.
\begin{equation}
    \label{eq:theta_it_est}\hat \theta^{(X+1)}_{\alpha} = \sum_{\alpha'} \frac{1}{2} F^{-1}_{\alpha\alpha'}(d_{\alpha'} - b_{\alpha'} - t_{\alpha'}),
\end{equation}
where
\begin{align}
    \label{eq:data_dn} d_{\alpha} &= \bm{x}^\mathrm{T} \mathbf{C}^{-1}\mathbf{Q}_{{\alpha}} \mathbf{C}^{-1} \bm{x}, \\
    \label{eq:noise_bn}b_{\alpha} &= \Tr(\mathbf{C}^{-1}\mathbf{Q}_{\alpha} \mathbf{C}^{-1}\mathbf{N}), \\
    \label{eq:signalfid_tn}t_{\alpha} &= \Tr(\mathbf{C}^{-1}\mathbf{Q}_{\alpha} \mathbf{C}^{-1}\mathbf{S}_{\mathrm{fid}}),
\end{align}
and the estimated Fisher matrix is
\begin{equation}
    \label{eq:fisher_matrix}F_{\alpha\alpha'} \equiv \left\langle \partderiv{^2 \lhf}{\theta_{\alpha} \partial \theta_{\alpha'}} \right\rangle = \frac{1}{2} \Tr(\mathbf{C}^{-1}\mathbf{Q}_{\alpha} \mathbf{C}^{-1} \mathbf{Q}_{\alpha'} ).
\end{equation}
The covariance matrices in the right hand side of equation~\ref{eq:theta_it_est} are computed using parameters from the previous iteration $\theta_{\alpha}^{(X)}$.

\subsection{Ly-$\alpha$ Forest Specifics}
In the Ly-$\alpha$ forest analysis, our data set $\bm x$ is a collection of pixels representing the normalized flux fluctuations $\bm{\delta}_F$. Assuming different quasar spectra are uncorrelated, the covariance matrix becomes block diagonal, where only the correlations within a spectrum are non-zero\footnote{The 3D analysis inherently needs these correlations between spectra, so this would not hold true.}. For example, stacking three quasar spectra $\bm{\delta}_{1,2,3}$ yields:
\begin{gather}
    \bm{\delta}_F = \left( \begin{array}{c}
         \bm{\delta}_1  \\
         \bm{\delta}_2 \\
         \bm{\delta}_3
    \end{array}\right) 
    \rightarrow
    \mathbf{C} = \left(\begin{array}{ccc}
        \mathbf{C}_1 & 0 & 0\\
        0 & \mathbf{C}_2 & 0 \\
        0 & 0 & \mathbf{C}_3
    \end{array} \right).
\end{gather}
This block diagonal structure simplifies equation~(\ref{eq:theta_it_est}) as well: The Fisher matrix $F_{\alpha\alpha'}$ and the expression in parentheses can be computed for each quasar, then accumulated, i.e. $\mathbf{F}=\sum_q\mathbf{F}_{q}$ etc. 

The correlation (signal) between pixels depends on their velocity separation, underlying power spectrum and spectrograph window function. We convert a pixel's wavelength to velocity using logarithmic spacing.
\begin{align}
    v_i &= c \ln (\lambda_i/\bar \lambda) \\
    z_i &= (1+ \bar z) \mathrm{e}^{v_i/c} - 1,
\end{align}
where $\bar \lambda$ and $\bar z$ are the median wavelength and the median redshift of the spectrum respectively\footnote{The pivot point does not matter as long as $v=c\ln\lambda$, and therefore the Ly-$\alpha$ rest-frame wavelength can also be used instead of the median wavelength of the spectrum.}. In general, the signal is multiplied with a resolution matrix $\mathbf{R}$, such that $\bm{\tilde s} = \mathbf{R}\bm{s}$ and therefore $\mathbf{\tilde S} = \mathbf{R}\mathbf{S}\mathbf{R}^{\mathrm{T}}$ \citep{bolton_spectro-perfectionism_2010}. DESI will provide this resolution matrix in its pipeline, which will be one of QMLE's strengths for future analyses. For the rest of the paper, we make the approximation that the resolution does not change with wavelength. Then, the signal becomes the correlation function convolved with the spectrograph resolution, which is the power spectrum multiplied with the spectrograph window function $W(k)$ in Fourier space.
\begin{equation}
    S_{ij}^{\mathrm{fid}} = \int_0^\infty \frac{dk}{\pi} \cos(k v_{ij}) W^2(k) P_{\mathrm{fid}}(k, z_{ij}),
\end{equation}
where $v_{ij}\equiv v_i - v_j$ and $1 + z_{ij} \equiv \sqrt{(1+z_i)(1+z_j)}$. The spectrograph window function is given by
\begin{equation}
    \label{eq:spectrograph_window}W(k) = e^{-k^2R^2/2} \text{sinc}(k\triangle v/2),
\end{equation}
where $R$ is the 1$\sigma$ resolution and $\triangle v$ is the pixel width, both in velocity units.

We now move on to defining power spectrum related specifics from data related specifics. In theory, $w_{\alpha}(k,z)$ can be any function, but we adopt top-hat $k$ bands with $k_n$ as bin edges and linear interpolation for $z$ bins with $z_m$ as bin centres: $w_{(mn)}(k,z) = H(k-k_n)H(k_{n+1}-k)I_m(z)$, where $\alpha \equiv (mn)$, $H(x)$ is the Heaviside step function and $I_m(z)$ is the interpolation kernel. We chose the linear interpolation for its smoothness over top hats. However, this mandates distributing pixel pairs into two redshift bins. One can imagine using higher order terms (such as cubic interpolation) to make the function smoother, and hence more accurate, but this will obviously make the calculation more complex. In the end, the linear interpolation is a good compromise between accuracy and complexity.
\begin{equation}
\label{eq:interp_kernel}
    I_{m}(z) =\begin{cases}
			\frac{z - z_{m-1}}{z_{m} - z_{m-1}} &,  z_{m-1} < z < z_{m} \\
			\frac{z_{m+1} - z}{z_{m+1} - z_{m}} &, z_{m} < z < z_{m+1} \\
			0  &, \text{otherwise}
	\end{cases}
\end{equation}
Note that this is 1 when $z=z_m$ and 0 when $z=z_{m\pm1}$. The derivative matrix for redshift bin $m$ and wavenumber bin $n$ is then
\begin{equation}
    Q_{ij}^{(mn)} = I_m(z_{ij}) \int_{k_n}^{k_{n+1}} \frac{dk}{\pi} \cos(kv_{ij}) W^2(k).
\end{equation}
We compute these matrices for as many redshift bins as necessary for a given spectrum.

Finally, we assume that the noise of every pixel is independent. This results in a diagonal noise matrix with $N_{ii}=\sigma_i^2$, where $\sigma_i$ is the pipeline noise divided by the continuum and the mean normalized flux $\bar F(z)$.

\subsection{Continuum Limit}
\label{subsec:cont_limit}
It is sufficient to take the estimated power as $P_{\mathrm{est}}(k^{\mathrm{c}}_n, z_m) = P_{\mathrm{fid}}(k^{\mathrm{c}}_n, z_m) + \theta_{(n,m)}$ in our analysis, where $k^{\mathrm{c}}_n$ is the bin centre. However, in order to further improve our intuition, let us discuss what equation~(\ref{eq:theta_it_est}) constructs in detail. First, we should differentiate between the underlying true power $P_{\mathrm{true}}$ of data and the fiducial power $P_{\mathrm{fid}}$ of the estimator; these two are not necessarily the same. For simplicity, let us ignore redshift dependence, spectrograph resolution and noise, and adopt band powers for $w_n(k)=H(k-k_n)H(k_{n+1}-k)$. In the continuum limit, matrix multiplications can be converted into integrals. Then, equation~(\ref{eq:theta_it_est}) becomes the following at iteration $X$:
\begin{align}
    \label{eq:cont_limit}\theta^{(X+1)}_n &= \int_{k_n}^{k_{n+1}}
    dk \; \gamma^{(X)}_n(k) [P_{\mathrm{true}}(k) - P_{\mathrm{fid}}(k)], \\
    \gamma^{(X)}_n(k) &= \frac{1}{P^2_{(X)}(k)}\left[\int_{k_n}^{k_{n+1}}
    \frac{dk}{P^2_{(X)}(k)} \right]^{-1},
\end{align}
where $P_{(X)}$ is exactly the fiducial power at the first iteration, and approaches to the true underlying power $P_{\mathrm{true}}$ by the last iteration. Thus, this estimator gives us an inverse variance weighted average of the residuals at each iteration. As these residuals gets smaller, the effect of averaging gets smaller as well.

\section{Implementation}
\label{sec:implementation}
QMLE implementation presents challenges in memory, CPU time, numerical stability and confident validation of the results. In this section, we clarify our implementation decisions in order to overcome such challenges. 

We will refer \citet{palanque-delabrouille_one-dimensional_2013} as \citetalias{palanque-delabrouille_one-dimensional_2013}, \citet{walther_new_2017} as \citetalias{walther_new_2017} and \citet{mcdonald_ly$upalpha$_2006} as \citetalias{mcdonald_ly$upalpha$_2006} from now on.

\subsection{Algorithm}
\label{subsec:algorithm}
At every iteration, the algorithm for each spectrum is as follows:
\begin{enumerate}
    \item\label{step:cov} Compute the covariance matrix using the previous iteration's $\hat{\bm{\theta}}$ estimates, then invert the covariance matrix. Note that the fiducial signal matrix stays fixed.  
    
    \item Compute weighted data vector $\bm{y} = \mathbf{C}^{-1} \bm{\delta}_F$, then $d_{\alpha} = \bm{y}^\mathrm{T} \mathbf{Q}_{\alpha} \bm{y}$.
    
    \item Compute $\mathbf{C}^{-1}\mathbf{Q}_{\alpha} \mathbf{C}^{-1}$, then $b_{\alpha}$ and $t_{\alpha}$ using equations~(\ref{eq:noise_bn}) and~(\ref{eq:signalfid_tn}).
    
    \item Compute Fisher matrix using equation~(\ref{eq:fisher_matrix}). Note that this needs to consider all redshift bins that the spectrum spans.
\end{enumerate}
We then sum $d_{\alpha}$, $b_{\alpha}$, $t_{\alpha}$ and $F_{{\alpha}{\alpha}'}$ of every quasar. Finally, we invert $\mathbf{F}$ and find $\hat{\bm{\theta}}$ estimates using equation~(\ref{eq:theta_it_est}). We check for convergence by comparing these results to the previous iteration using the expressions in section~\ref{subsec:convergence}.

One can bootstrap QMLE results easily by saving $F_{{\alpha}{\alpha}'}$ and $(d_{\alpha}-b_{\alpha}-t_{\alpha})$ of each spectrum to a file. Since each spectrum is treated independently, there is no need to recompute these for every realization. One can generate as many bootstrap realizations as needed on spectrum level, and then simply add these saved quantities with repetition to find a bootstrapped estimate. This treatment is exact at the first iteration, and should be a good approximation at convergence.

We find that using a smooth weighted spline in step~\ref{step:cov} makes the algorithm numerically stable. This smoothing spline is performed on $(k, P=P_{\mathrm{fid}}+\theta)$, and it can be more reliable if performed on $(\ln k, \ln P)$ while non-positive values are removed.

This algorithm can be implemented using only three matrices for each quasar: One holds the covariance matrix and its inverse, and the other two temporarily hold the derivative and fiducial signal matrices. Holding all matrices in memory for a quasar at a time can improve computation time when memory is the lesser concern, which we find possible in our tests.

This method demands substantial CPU time, and therefore some optimisation mechanisms are noteworthy. First, instead of integrating $\mathbf{S}_{\mathrm{fid}}$ and $\mathbf{Q}_{\alpha}$ when needed, we create lookup tables once and interpolate. Moreover, every spectrum can be computed in parallel as they are assumed independent. Each CPU should have near-equal workload for efficient parallelization. Matrix operations scale as $\mathcal{O}(N_D^3)$, where $N_D$ is the number of pixels in a spectrum. Let us define $N_B$ as the number of total bins to which a spectrum contributes. Then, the Fisher matrix calculation will require $N_B(N_B+1)/2$ matrix multiplications for that spectrum. Hence, we use $T_{\mathrm{cpu}}=N_D^3 N_B(N_B+1)$ as an estimate for computation time, and distribute spectra accordingly. 

We use \textsc{gsl}\footnote{\url{https://www.gnu.org/software/gsl/}} for interpolation, integration, matrix inversion; Intel's \textsc{mkl} library\footnote{\url{https://software.intel.com/en-us/mkl}} for matrix multiplication; and we compile with \textsc{open-mpi}\footnote{\url{https://www.open-mpi.org}} for parallelization. Smooth weighted spline is constructed using \textsc{scipy}\footnote{\url{https://www.scipy.org}}.

\begin{table}
\centering
\caption{Top: \citetalias{palanque-delabrouille_one-dimensional_2013} BOSS likelihood fitting parameters. Bottom: This work fitting  combined \citetalias{palanque-delabrouille_one-dimensional_2013} and \citetalias{walther_new_2017} data with Lorentzian decay added.}
	\begin{tabular}{|c|c|c|c|c|c|}
	\hline
	$A$ & $n$ & $\alpha$ & $B$ & $\beta$ & $k_1$ [km s$^{-1}$]\\
   	\hline
	$0.06$ & $-2.55$ & $-0.10$ & $3.55$ & $-0.28$ & --\\
	0.066 & $-2.685$ & $-0.22$ & 3.59 & $-0.18$ & 0.53  \\ 
    \hline
	\end{tabular}
	\label{table:pdw_fit_params}
\end{table}

\subsection{Fiducial Power Spectrum}
We exploit a baseline estimate of power spectrum in our analysis as discussed in section~\ref{sec:theory}. \citetalias{palanque-delabrouille_one-dimensional_2013} provides a fitting function with best-fitted parameters. We further modify their fitting function with a Lorentzian decay:
\begin{equation}
    \label{eq:pd13_fitting_fn}\frac{kP(k, z)}{\pi} = A \frac{(k/k_0)^{3 +n + \alpha\ln k/k_0}}{1+(k/k_1)^2} \left(\frac{1+z}{1+z_{0}}\right)^{B + \beta\ln k/k_0},
\end{equation}
where $k_{0} = 0.009$ km s$^{-1}$ and $z_{0}=3.0$. However, \citetalias{palanque-delabrouille_one-dimensional_2013} measures power spectrum up to $0.02$ km s$^{-1}$, so their parameters are not valid on small scales. We combine \citetalias{walther_new_2017}'s power spectrum estimates, and fit the resulting data set. Our modification reduces $\chi^2_{\nu}$ from 16.5 to 5.6. Even though the fit should not be used for scientific purposes, it should be sufficient for a baseline estimate.

Our implementation is also equipped with taking a tabulated fiducial power as input. This feature gives greater freedom in the choice of fiducial. We use this to eliminate any discord between synthetic data and the estimator; and to investigate different choices as fiducial.

\subsection{Convergence}
\label{subsec:convergence}
We choose a convergence criterion that summarizes the overall fluctuations between iterations. For this purpose, we calculate the weighted average of the changes between iterations using estimated Gaussian errors. In other words, we define convergence when
\begin{equation}
    \label{eq:convergence_chi}\triangle \chi = \sqrt{ \frac{1}{N}\sum_{\alpha} \frac{(\triangle \hat \theta_{\alpha})^2}{F^{-1}_{{\alpha}{\alpha}}}}
\end{equation}
or
\begin{equation}
    \triangle \chi_F = \sqrt{\frac{1}{N}(\bm{\Delta \hat \theta})^{\mathrm{T}} \mathbf{F} (\bm{ \Delta \hat \theta})}
\end{equation}
becomes smaller than $\chi_c = 0.01$, where $\triangle \hat\theta_{\alpha} \equiv \hat\theta_{\alpha}^{(X+1)} - \hat\theta_{\alpha}^{(X)}$ and $N$ is the total number of bins. Both expressions return close values and reach convergence at the same iteration in almost all cases.

\begin{figure}
    \centering
    \includegraphics[width=\columnwidth]{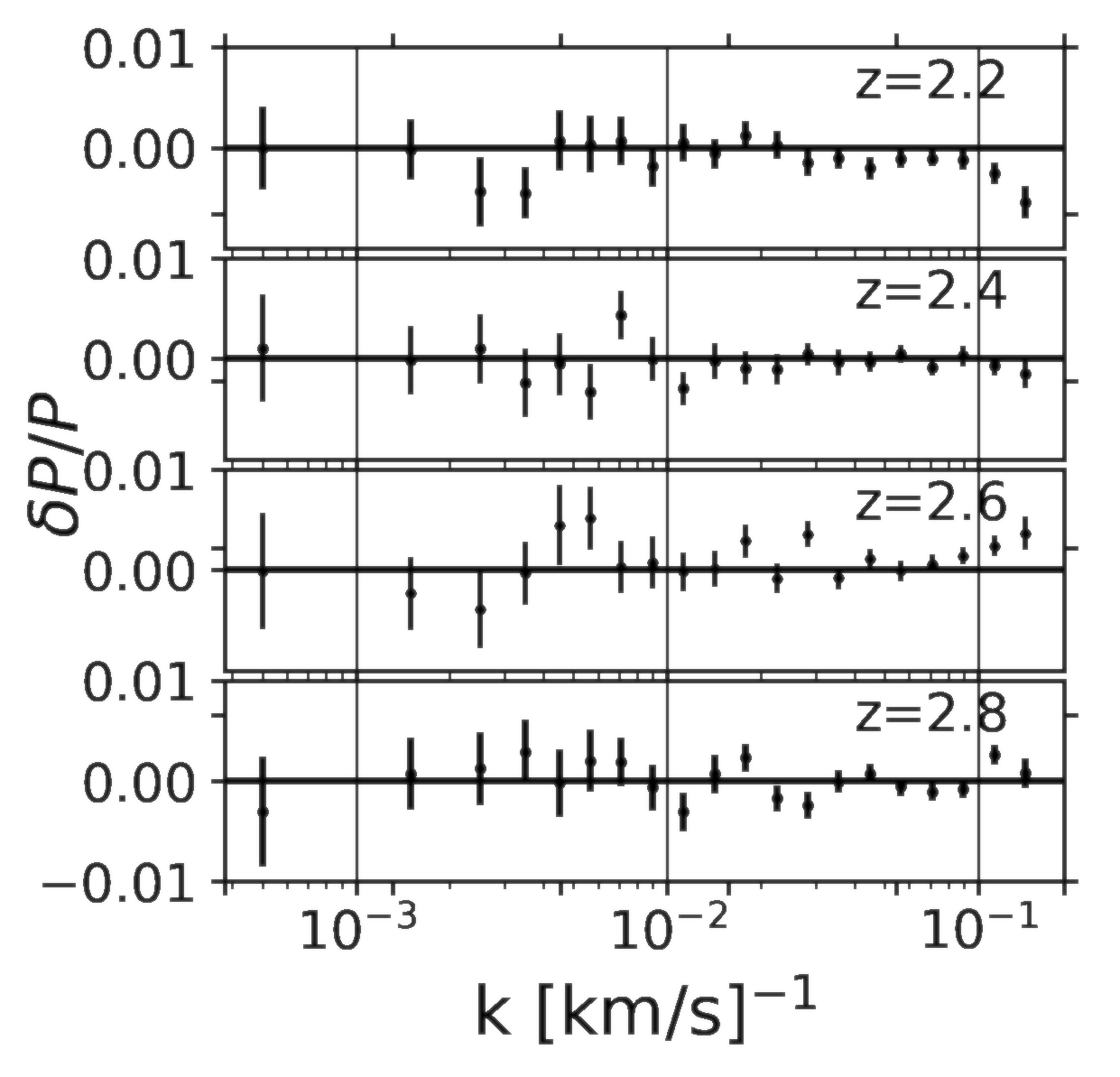}
    \caption{The relative error in our measurements from 100 log-normal catalogues where each has 1000 spectra in every redshift bin. Our method reaches sub-percent level precision in the absence of any systematic. Bins go up to the Nyquist frequency.}
    \label{fig:lognormal1000}
\end{figure}

\begin{figure*}
    \includegraphics[width=0.32\linewidth]{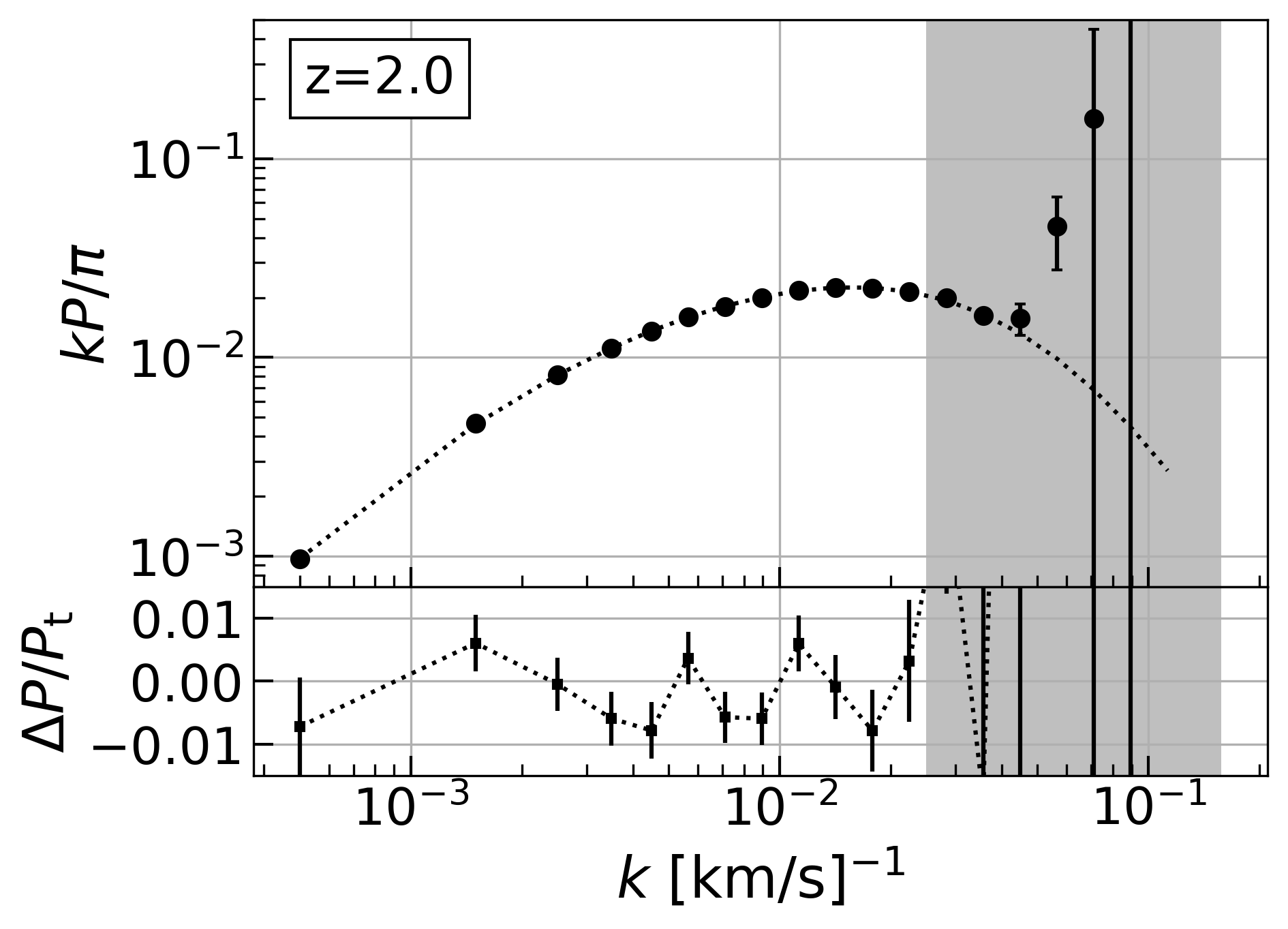}
    \includegraphics[width=0.32\linewidth]{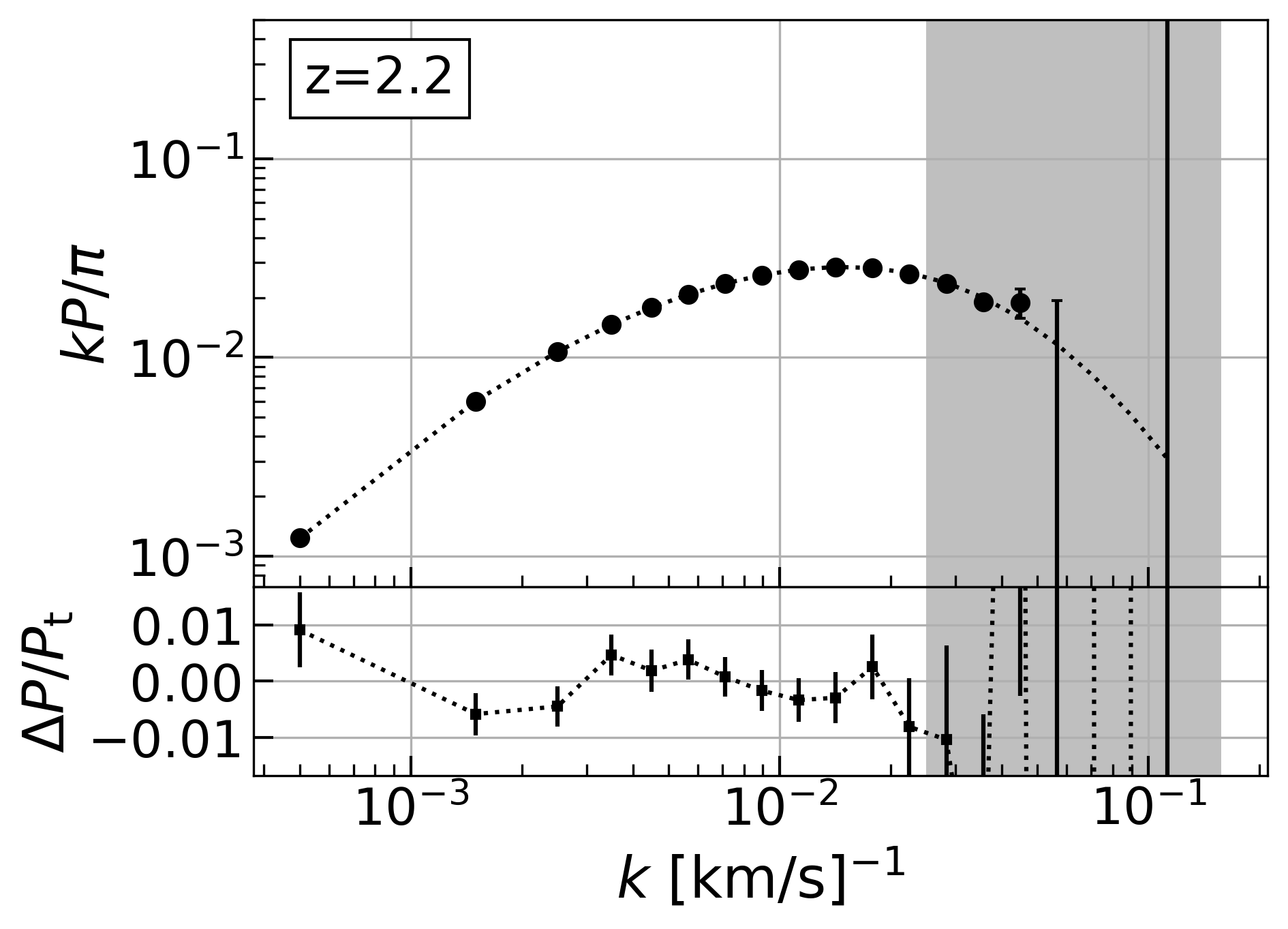}
    \includegraphics[width=0.32\linewidth]{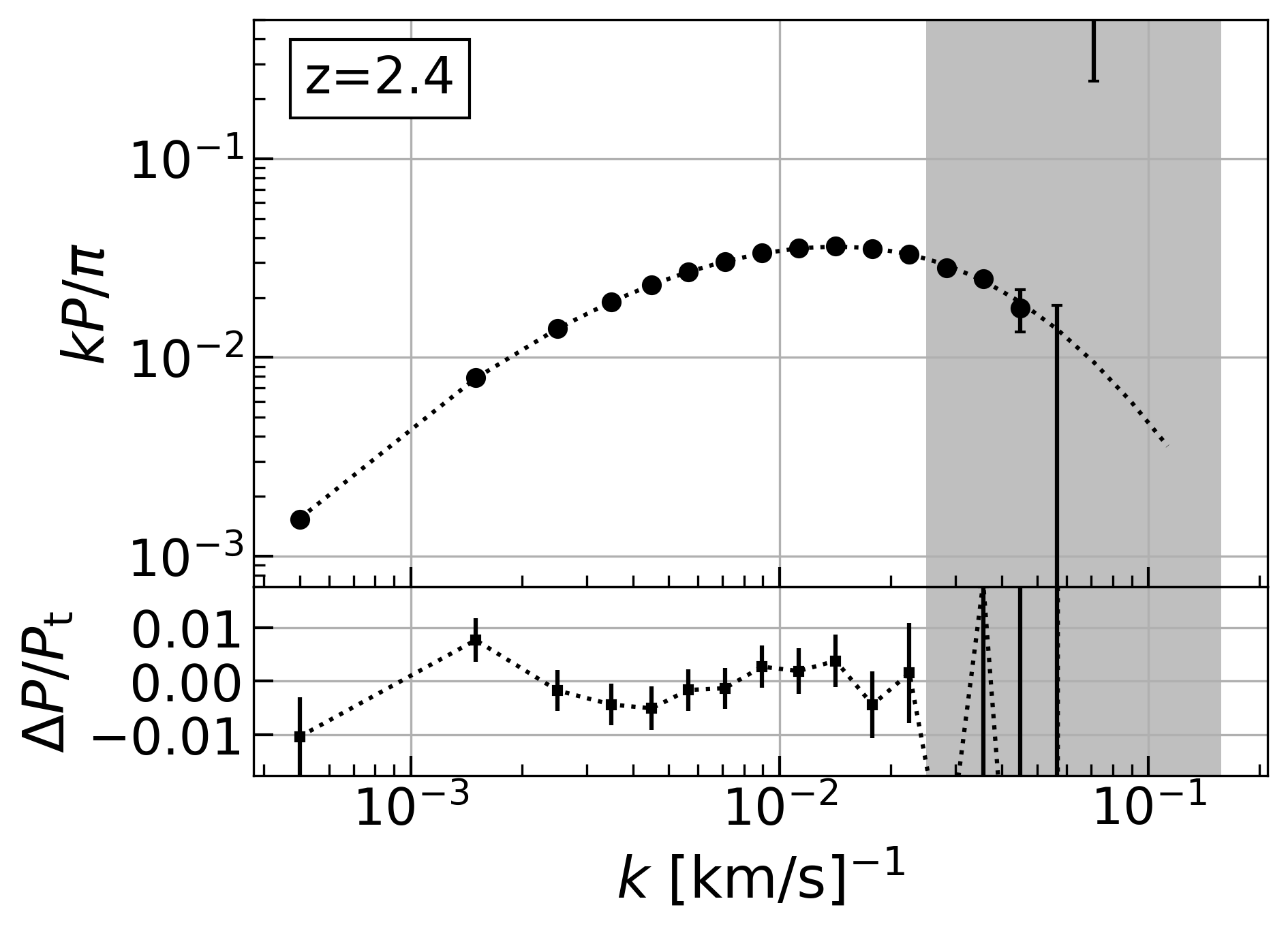}\\
    \includegraphics[width=0.32\linewidth]{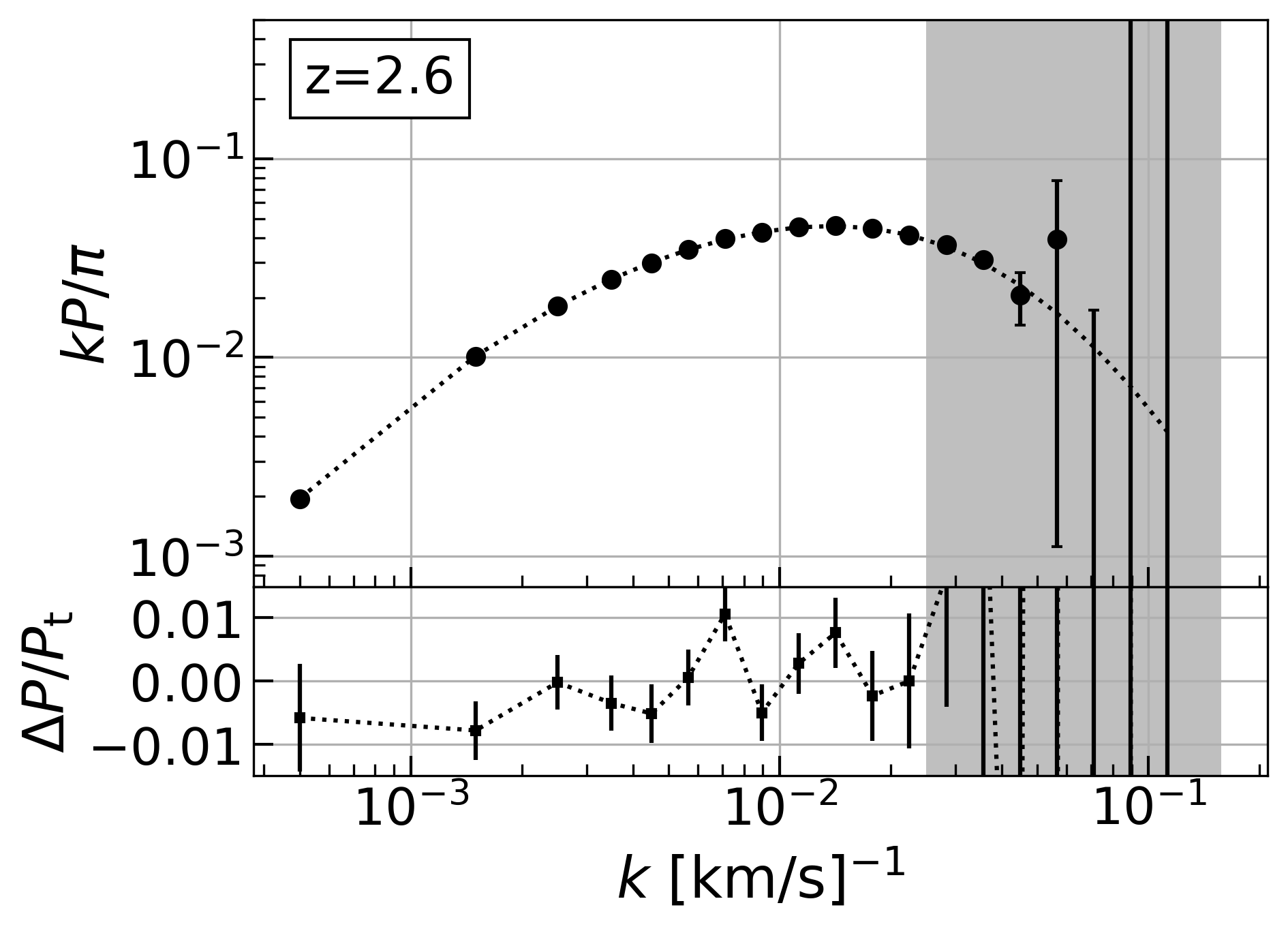}
    \includegraphics[width=0.32\linewidth]{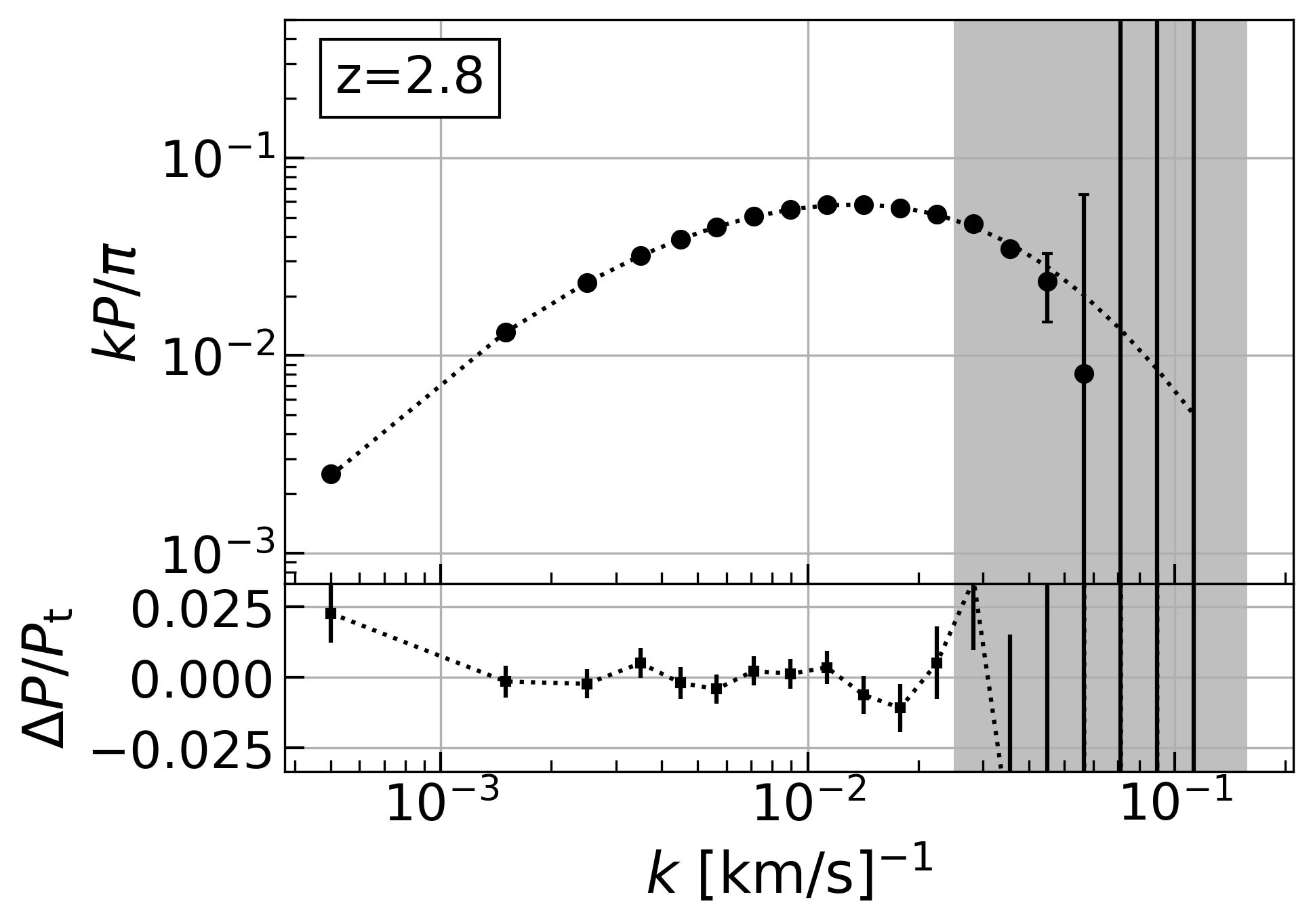}
    \includegraphics[width=0.32\linewidth]{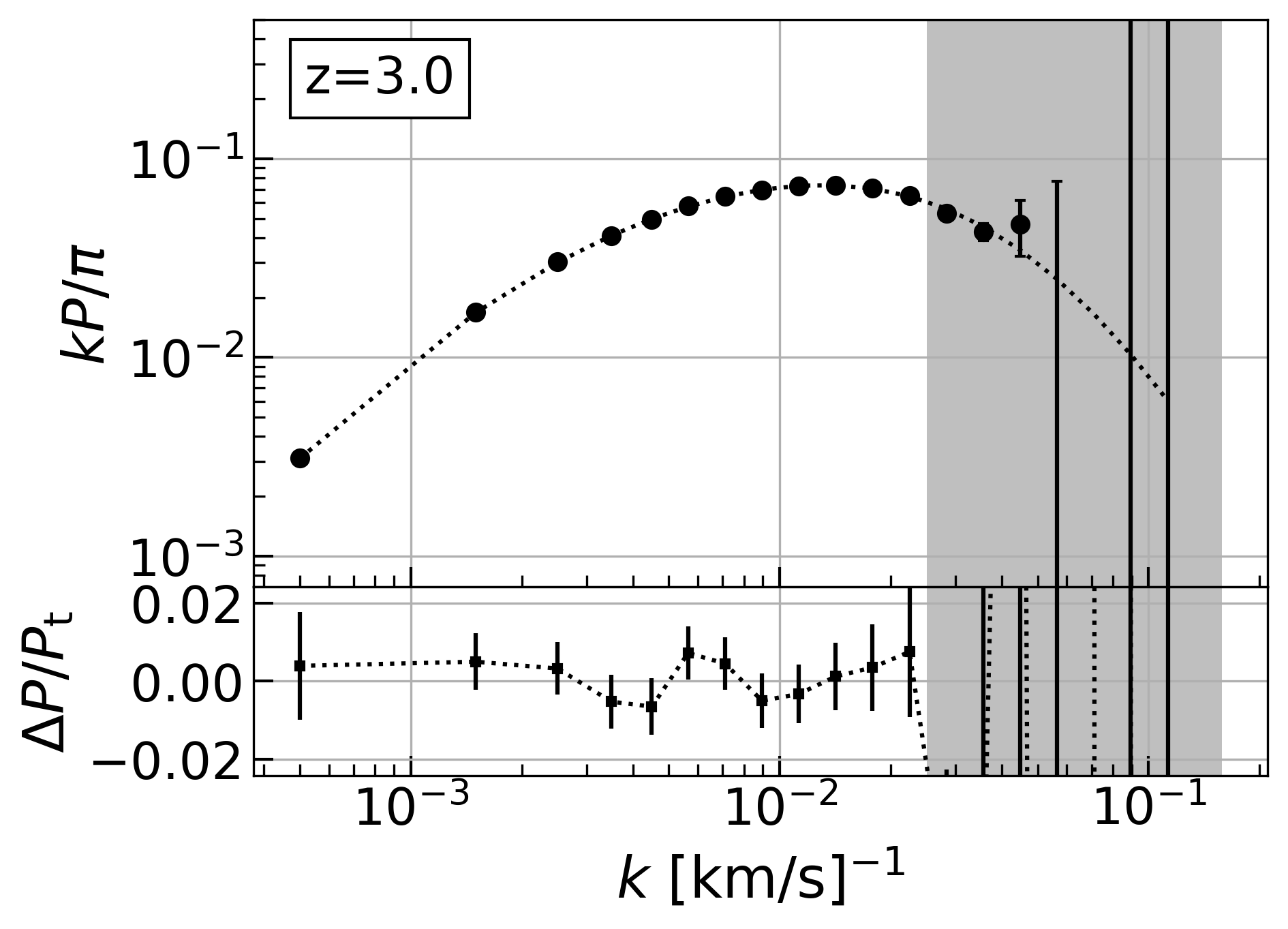} \\
    \includegraphics[width=0.32\linewidth]{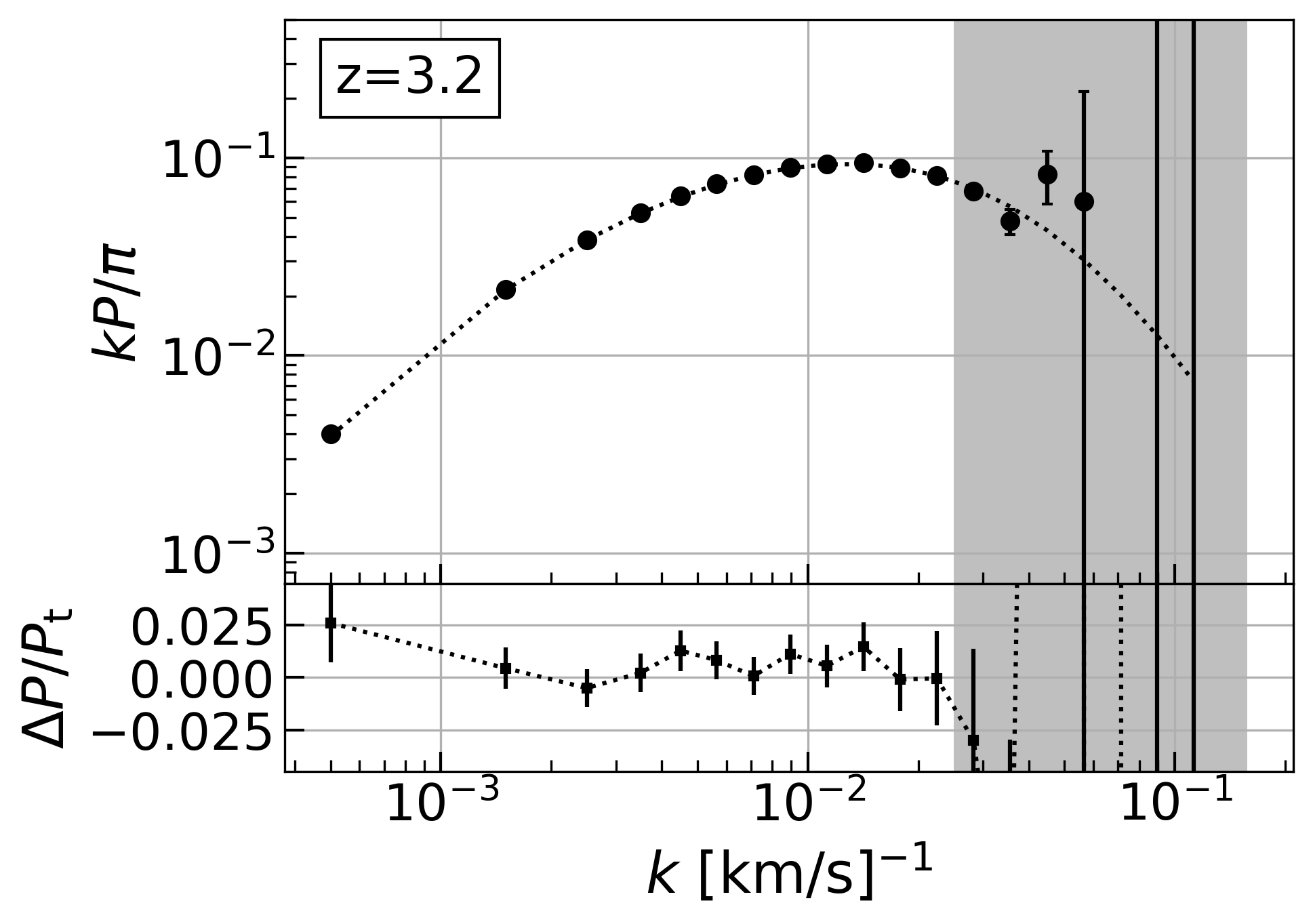}
    \includegraphics[width=0.32\linewidth]{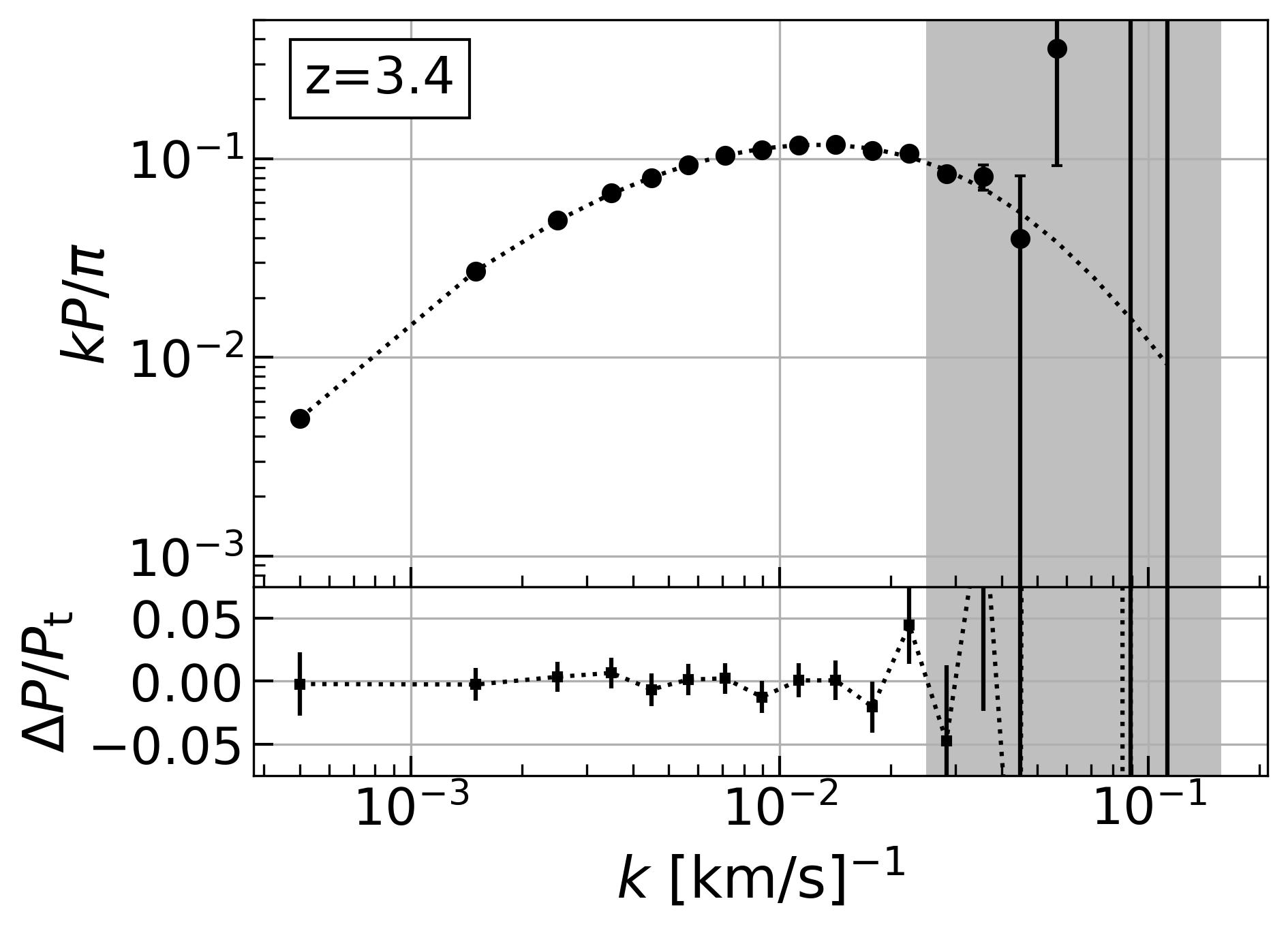}
    \includegraphics[width=0.32\linewidth]{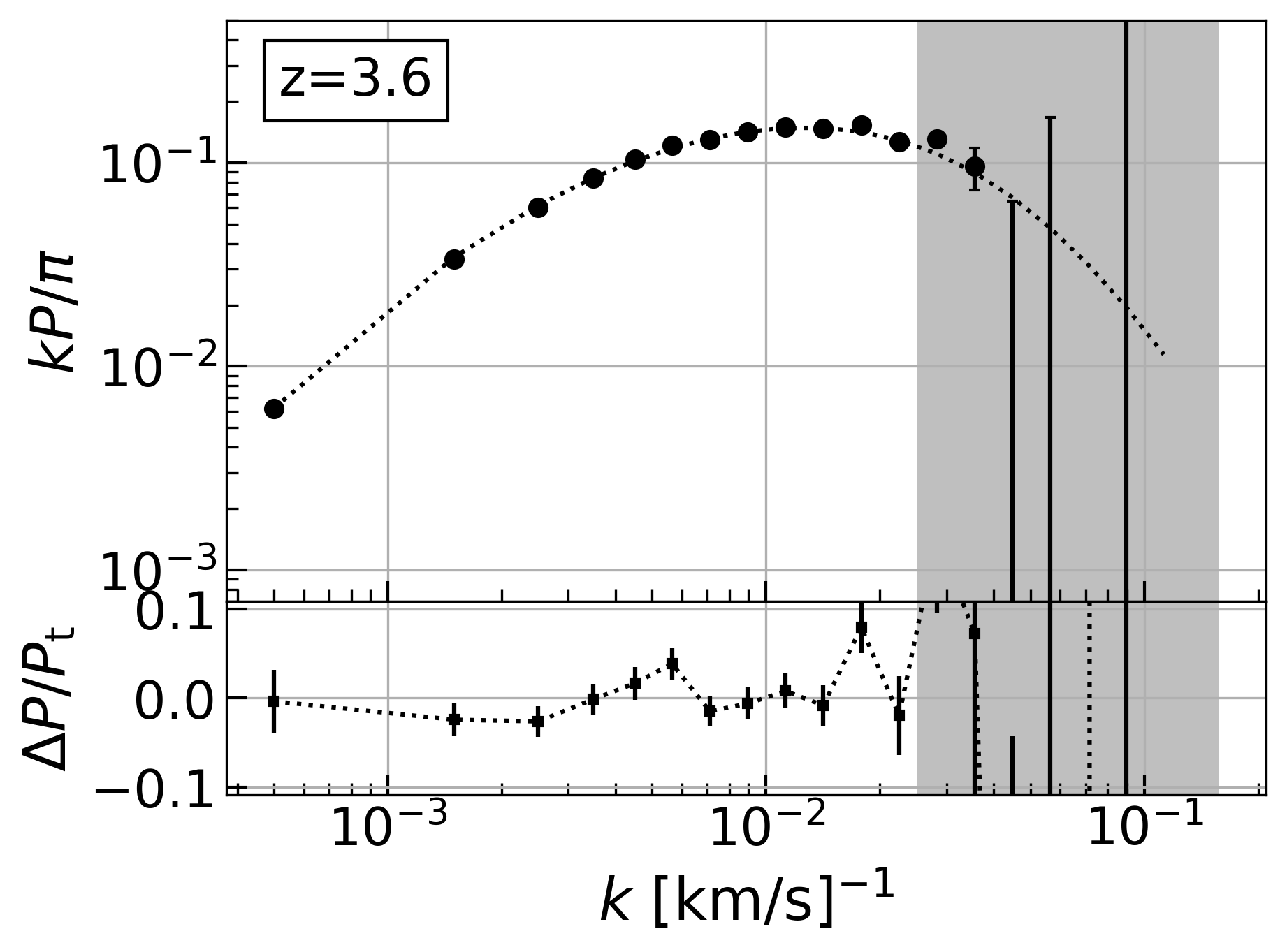} \\
    \includegraphics[width=0.32\linewidth]{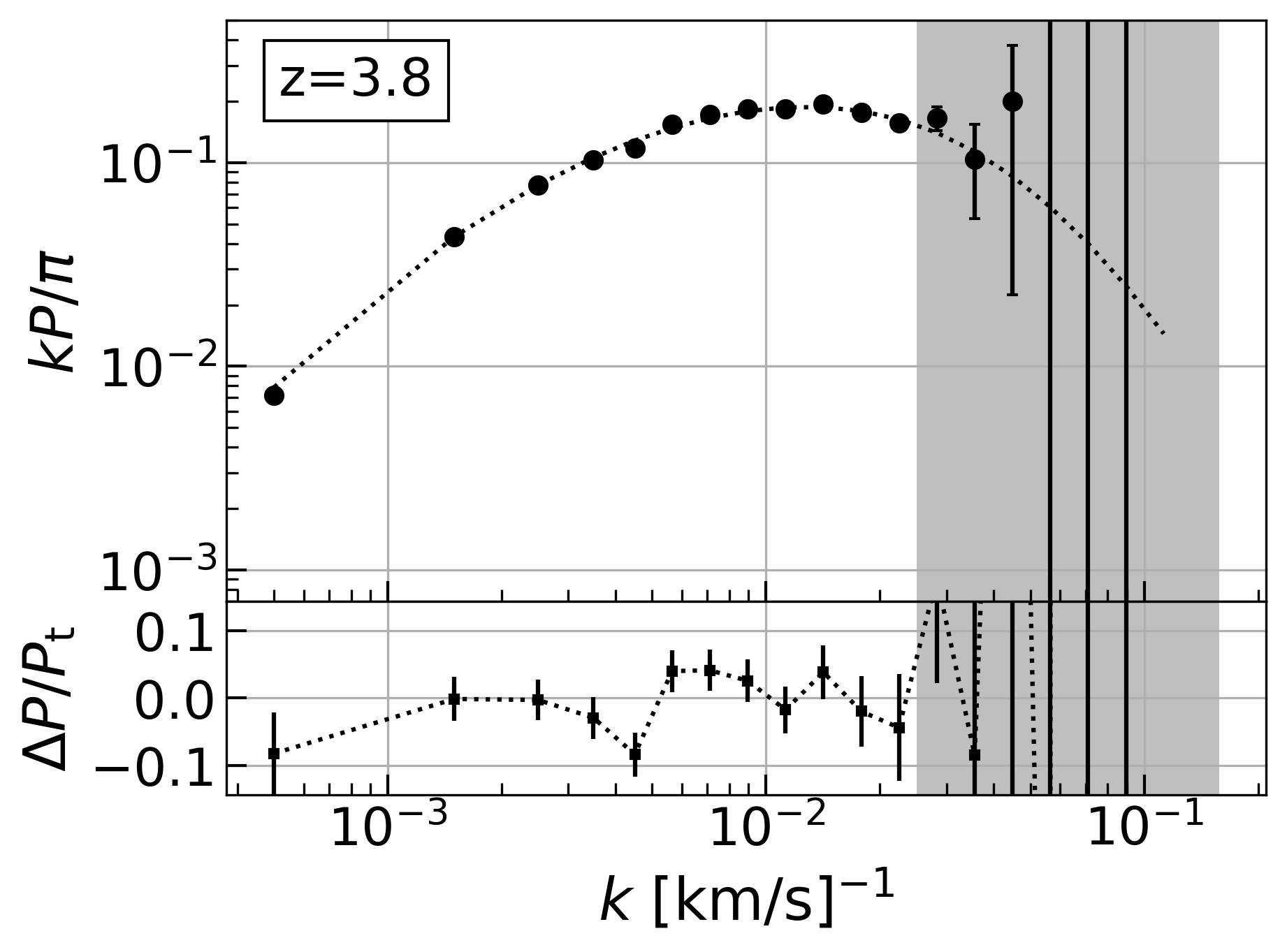}
    \includegraphics[width=0.32\linewidth]{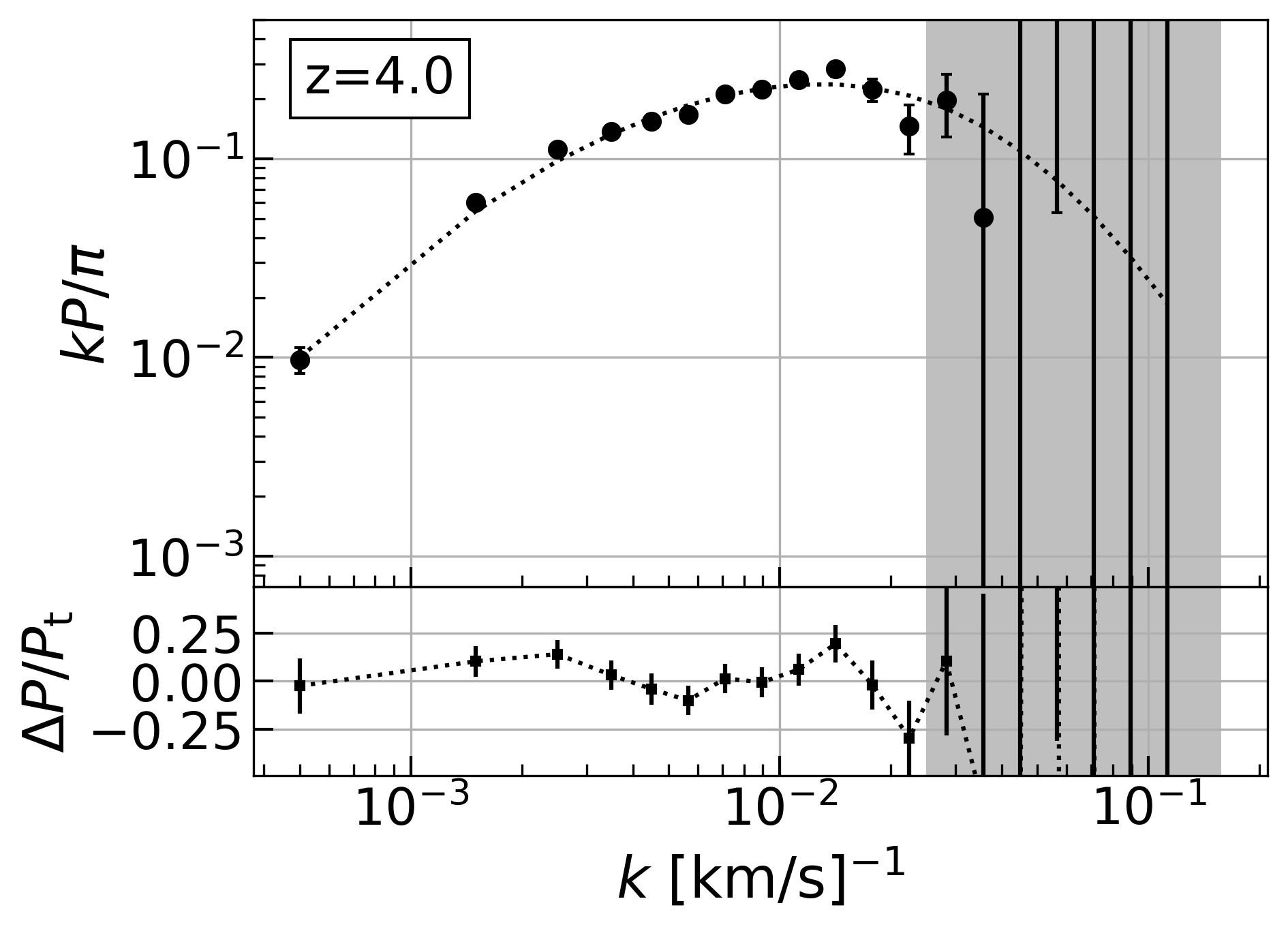}
    \includegraphics[width=0.32\linewidth]{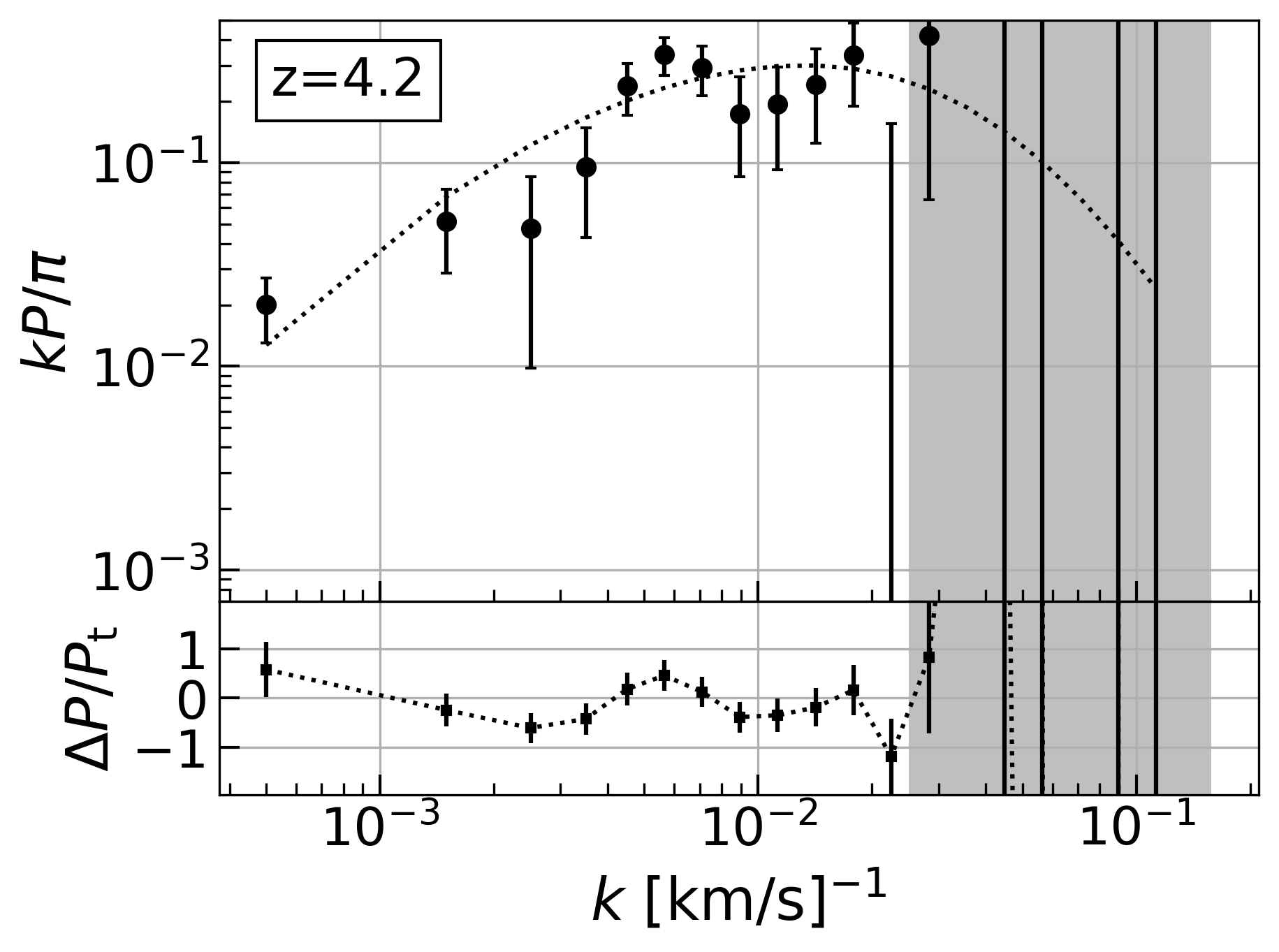}
    \caption{Power spectrum estimates from 800,000 DESI-lite spectra. Error bars from the diagonal of the inverse Fisher matrix. Dotted line in each upper panel is the true power. Noise begins to dominate in grey shaded area, $k > k_N = 0.025$ s km$^{-1}$. The estimates reach percent-level accuracy in lower redshift bins. However, last redshift bins are poorly estimated due to significant decline in statistics. The $\chi^2$ calculated from diagonal (all) elements is 240 (248) for 228 degrees of freedom.}
    \label{fig:desi800k}
\end{figure*}

\subsection{Validation}
We use log-normal mocks for validation. These semi-realistic spectra are crucial to examine the accuracy, precision and efficiency of our method. 

We generate log-normal mocks by using a modified version of \citetalias{mcdonald_ly$upalpha$_2006}. These realizations approximately produce theoretically expected mean flux redshift evolution \citep{faucher-giguere_direct_2008, becker_refined_2013} and power spectra  similar to \citetalias{palanque-delabrouille_one-dimensional_2013} and \citetalias{walther_new_2017}.

\begin{enumerate}
    \item Generate a long high-resolution Gaussian random grid with equal spacing in velocity $v$, zero mean and unit variance.
    
    \item FFT this grid and multiply with $\sqrt{P(k)/dv}$ to obtain $\tilde \delta_{b}(k)$, where $dv$ is the grid spacing in velocity units and the power spectrum is
    \begin{equation}
        P(k) = \frac{(k/k_0)^{n-\alpha \ln(k/k_0)}}{1 + (k/k_1)^\gamma},
    \end{equation}
    where $k_0=0.001$ s km$^{-1}$, $k_1 = 0.04$ s km$^{-1}$, $n=0.5$, $\alpha=0.26$ and $\gamma=1.8$. Inverse FFT and save the variance of this grid $\sigma^2$. This is a crude Gaussian base for baryon fluctuations $\delta_{b}(v)$ with defined power spectrum at $z_0=3$.
    
    \item Multiply with a redshift evolution factor $a(z)$. Such that $\delta_b(z)= a(z) \delta_{b}$ and $\sigma^2(z)=a^2(z)\sigma^2$.
    \begin{equation}
        a^2(z) = 58.6 \left(\frac{1+z}{1+z_0}\right)^{-2.82}
    \end{equation}
    
    \item Apply a squared log-normal transformation to approximate the non-linear and non-Gaussian H\textsc{i} column density field.
    \begin{equation}
        n(z) = e^{2\delta_b(z) - \sigma^2(z)}
    \end{equation}
    
    \item Transform this to optical depth $\tau$ by multiplying with another redshift dependent function.
    \begin{equation}
        \tau(z) = 0.55\left(\frac{1+z}{1+z_0}\right)^{5.1} n(z)
    \end{equation}
    
    \item Finally, the flux is $F(z) = e^{-\tau(z)}$.
    
    \item Smoothing $F(z)$ with a Gaussian kernel and re-sampling it onto the observed wavelength grid will result in a spectrograph function in equation~(\ref{eq:spectrograph_window}).
\end{enumerate}

\begin{figure}
    \centering
    \includegraphics[width=0.8\linewidth]{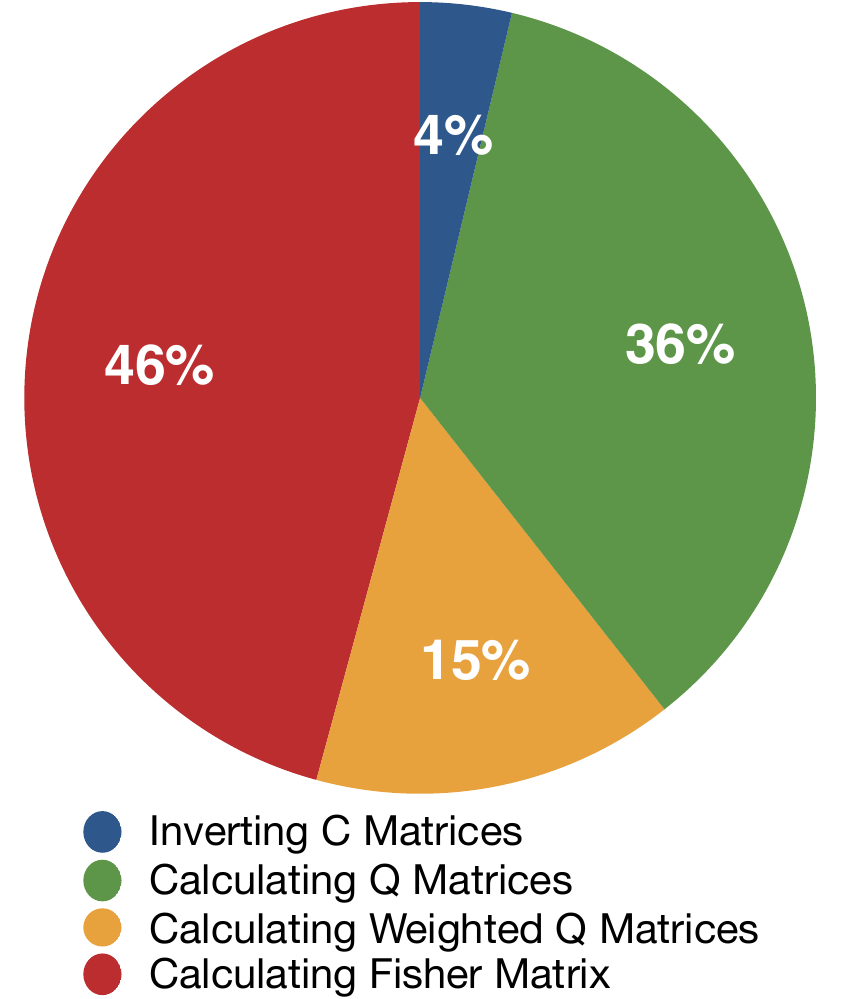}
    \caption{Percentage of time spent in different steps for 800,000 DESI-lite spectra. The weighted Q matrices are $\mathbf{C}^{-1}\mathbf{Q}_{\alpha}\mathbf{C}^{-1}$. Fisher matrix calculation consumes the most time, which also requires calculating $\mathbf{Q}_{\alpha}$ matrices as prerequisite. The time for remaining steps are insignificant.}
    \label{fig:time_stat}
\end{figure}

The mean flux and power spectrum of these mocks can be analytically computed. Using one-point probability of the base Gaussian random field $\delta$, we can write down the following integral for the mean flux:
\begin{align}
    \bar F(z) &= \frac{1}{\sigma\sqrt{2\pi}} \int_{-\infty}^{\infty} \exp\left[- \frac{\delta^2}{2\sigma^2} - x(z) e^{2a(z)\delta}\right] d\delta,
    \label{eq:theo_mean_flux}
\end{align}
where we have defined
\begin{equation}
    x(z) \equiv 0.55 \left(\frac{1+z}{1+z_0}\right)^{5.1} e^{- a^2(z)\sigma^2}.
\end{equation}
Although we have found a closed form using saddle point integration (see Appendix~\ref{app:saddle}), it deviates from the truth for $z\lesssim 3$ up to 4\%. It is more accurate to just integrate this expression; and Gauss-Hermite quadrature provides a reliable fast integration. 

Power spectrum expression needs two-point probability. We start with flux fluctuations $\delta_F(v, z) = F(v, z)/\bar F(z) - 1$, then express the correlation function $\xi_F$ as an integral over two Gaussian random $\delta$s assuming all pixels are at the same redshift. We drop $z$ for clarity.
\begin{align}
    1+\xi_F(v_{ij}) &= \int \frac{e^{-\bm\delta^T \mathbf{C}^{-1} \bm\delta /2}}{2\pi \sqrt{\det \mathbf{C}}} \frac{F_iF_j}{\bar F^2} d\bm\delta,
\end{align}
where
\begin{align}
    \mathbf{C} &=\left(\begin{array}{cc}
        \sigma^2 & \xi_G(v_{ij}) \\
        \xi_G(v_{ij}) & \sigma^2
    \end{array}\right),
\end{align}
and $\xi_G(v)$ is the correlation function of the base Gaussian field. To convert this expression into Gauss-Hermite quadrature with two variables, we apply Cholesky decomposition $\mathbf{C}=\mathbf{L}\mathbf{L}^{\mathrm{T}}$ and transform $\bm\delta = \sqrt{2} \mathbf{L}\cdot\bm y$, where
\begin{align}
    \mathbf{L} &= \left( \begin{array}{cc}
        \sigma & 0 \\
        \frac{\xi_G(v)}{\sigma} & \sigma\sqrt{1 - \frac{\xi_G(v)^2}{\sigma^4}} 
    \end{array}\right)
\end{align}
Here is the final expression for completeness:
\begin{equation}
    1+\xi_F(v) = \int e^{-\bm y^2} \frac{1}{\pi \bar F^2(z)} \exp\left[-x(z) \left( e^{2a\delta_1}+ e^{2a\delta_2}\right) \right] d\bm y
\end{equation}

We perform an initial test in the absence of redshift distribution and resolution effects. We start with a long fine grid ($dv=0.4\bar{3}$ km s$^{-1}$, $N=2^{20}$), and generate mocks that are discretely distributed in redshift. For validation, we pick four redshifts (2.2, 2.4, 2.6, 2.8), and simulate 100 catalogues where each catalogue has 1000 spectra in every redshift bin. These spectra are exactly centred at these redshifts. We set the spectrograph resolution $R=71600$ to diminish its effect, limit the spectral length to half of the bin size $\Delta z = 0.1$, and finally re-sample to pixel size of $\Delta v = 20.8$ km s$^{-1}$. Figure~\ref{fig:lognormal1000} shows our results. Our method reaches sub-percent level accuracy in the absence of any systematic.

\begin{figure}
    \centering
    \includegraphics[width=\linewidth]{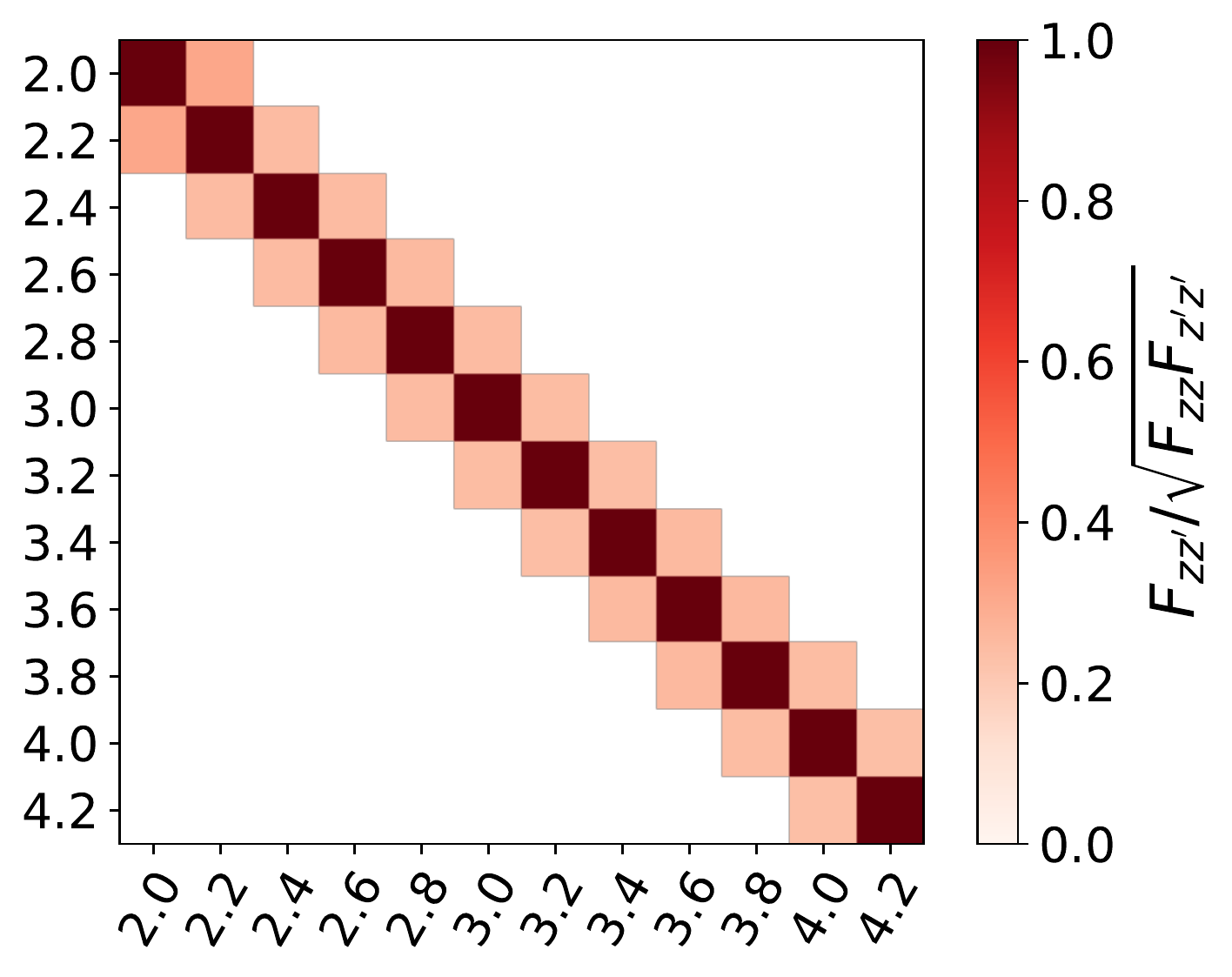}
    \caption{The normalized Fisher matrix between redshift bins for $k=0.0071$ s km$^{-1}$ bin. The off-diagonal terms are approximately $40\%$. Only neighbouring redshift bins dominate the Fisher matrix, since a pixel pair is split into two bins.}
    \label{fig:fisher_zbyz}
\end{figure}

\section{DESI-Lite Spectra}
\label{sec:desilite_spectra}
To test the feasibility of our method for future DESI spectra, we generate semi-realistic data set using the following simplifying assumptions: 
\begin{enumerate}
    \item We set the observed wavelength grid between 3600--9800 \AA. This means the closest forest pixel is at $z=1.96$.
    
    \item We create a logarithmically spaced wavelength grid with $\Delta v= 30$ km s$^{-1}$\footnote{DESI will use linear wavelength spacing in its pipeline, but QMLE does not require equal velocity spacing.}. Our grid spacing corresponds to a Nyquist frequency of $0.1$ s km$^{-1}$. 
    
    \item We assume a constant resolution power of $R=3200$ ($\approx 94$ km s$^{-1}$ FWHM in velocity units) for all spectra at all wavelengths.

    \item We add Gaussian random errors with $\sigma=0.7$ to $F$, which is approximately S/N= 2 \AA$^{-1}$.
    
    \item We set the minimum redshift of a Ly-$\alpha$ quasar to 2.1 and the maximum to 4.4, and pick random redshifts from a distribution $n(z)$ \citep{palanque-delabrouille_extended_2016, desi_collaboration_desi_2016}. We always limit the forest to [1050 \AA, 1180 \AA] range in quasar's rest frame.
    
    \item Finally, we assume DESI will observe 800,000 quasar spectra.
\end{enumerate} 

We choose 12 redshift bins between 2.0 and 4.2 with $\Delta z=0.2$. First 5 $k$ bins are linearly spaced with $\Delta k_{\mathrm{lin}}=0.001$ s km$^{-1}$, and the following 14 bins are logarithmically spaced with $\Delta k_{\mathrm{log}} = 0.1$. Hence, $0.0005 \text{ s km}^{-1} \leq k \leq 0.112 \text{ s km}^{-1}$.

Figure~\ref{fig:desi800k} shows power spectrum estimates for each redshift bin with error bars from the diagonal elements of the inverse Fisher matrix. The chi-square using the full Fisher matrix is $\chi^2/\nu=248/228$\footnote{We performed 9 independent runs and found $\chi^2$ fluctuating around 228 as expected. The value 248 is from our first run, and not a special case.}, which implies a valid agreement with the truth. We measure the power spectrum to sub-percent accuracy at lower redshifts, but the accuracy and precision get progressively worse towards high redshifts due to declining quasar numbers. We also expected noise and window function corrections to dominate at high $k$ as the noise power crosses the signal at $k_N \approx 0.025$ s km$^{-1}$\footnote{This $k_N$ roughly corresponds to $P_N/\sqrt{N_{\mathrm{qso}}}=P_{\mathrm{1D}}$, where the noise power is $P_N=\left(\sigma \Delta v/\bar F(z)\right)^2$ with $N_{\mathrm{qso}}=100,000$ for all redshift bins and $P_{\mathrm{1D}}$ is obtained analytically.}. This constitutes our foundation as validation of our method.

We would like to stress a subtle point here. First of all, the estimator constructs correct covariance matrices in the first iteration by using the true power as fiducial input. This means we also have the correct Fisher matrix and expect $\theta\approx 0$ after the iteration. However, the convergence criteria in equation~\ref{eq:convergence_chi} will still yield nearly 1 due to the statistical fluctuations of the power spectrum (note $\mathrm{Var}[\theta_{\alpha}] \sim F^{-1}_{\alpha\alpha}$). Therefore, when the estimator goes into the second iteration, it misidentifies these intrinsic fluctuations as corrections to the covariance matrices and readjusts the Fisher matrix. The smoothing spline ameliorates this digression, though imperfectly.

Fisher matrix calculation consumes the most time as it requires calculating $\mathbf{Q}_{\alpha}$ matrices and multiplying them $\mathcal{O}(N^2)$ many times. We identified that at least 46\% of the total time goes into the Fisher matrix (see figure~\ref{fig:time_stat}). Moreover, we also found that the Fisher matrix is a band matrix that is prominently tridiagonal for individual $k$ and $z$ bins. We normalize the Fisher matrix with respect to its diagonal elements for a clear representation in figure~\ref{fig:fisher_zbyz}, which is for $k=0.0071$ s km$^{-1}$ bin, but represents a typical redshift dependence\footnote{The lowest $k$ bin is weakly coupled $(\sim 1\%)$ to an additional redshift bin.}. This tridiagonal shape is due to keeping the full forest and distributing pixel pairs into two redshift bins. This structure weakly persists between $k$ bins for a given $z$ bin as well. Given Fisher matrix is the longest step, limiting its calculation to only these terms will speed up the estimation significantly by decreasing the number of matrix multiplications to $\mathcal{O}(N)$. We consider this optimisation scheme and how well it performs in section~\ref{sec:fisher_optm}. It is also worth pointing out that the Fisher matrix does not depend on the data, but only on the input power spectrum. Therefore, one could also choose a common Fisher matrix for analysis with many simulations.

We also computed the power spectrum covariance matrix with 5000 bootstrap realizations using the results of the first iteration as discussed in section~\ref{subsec:algorithm}. We found that the diagonals of the covariance matrix estimated from this bootstrap procedure agreed within percent level with the formal estimates from the Fisher matrix except for the high redshift bin where we do not have enough quasars to compute robust bootstrap errors. This bootstrapping procedure therefore provides a straightforward test of the assumptions underlying the error estimates from the Fisher matrix.

\begin{figure}
    \centering
    \includegraphics[width=\linewidth]{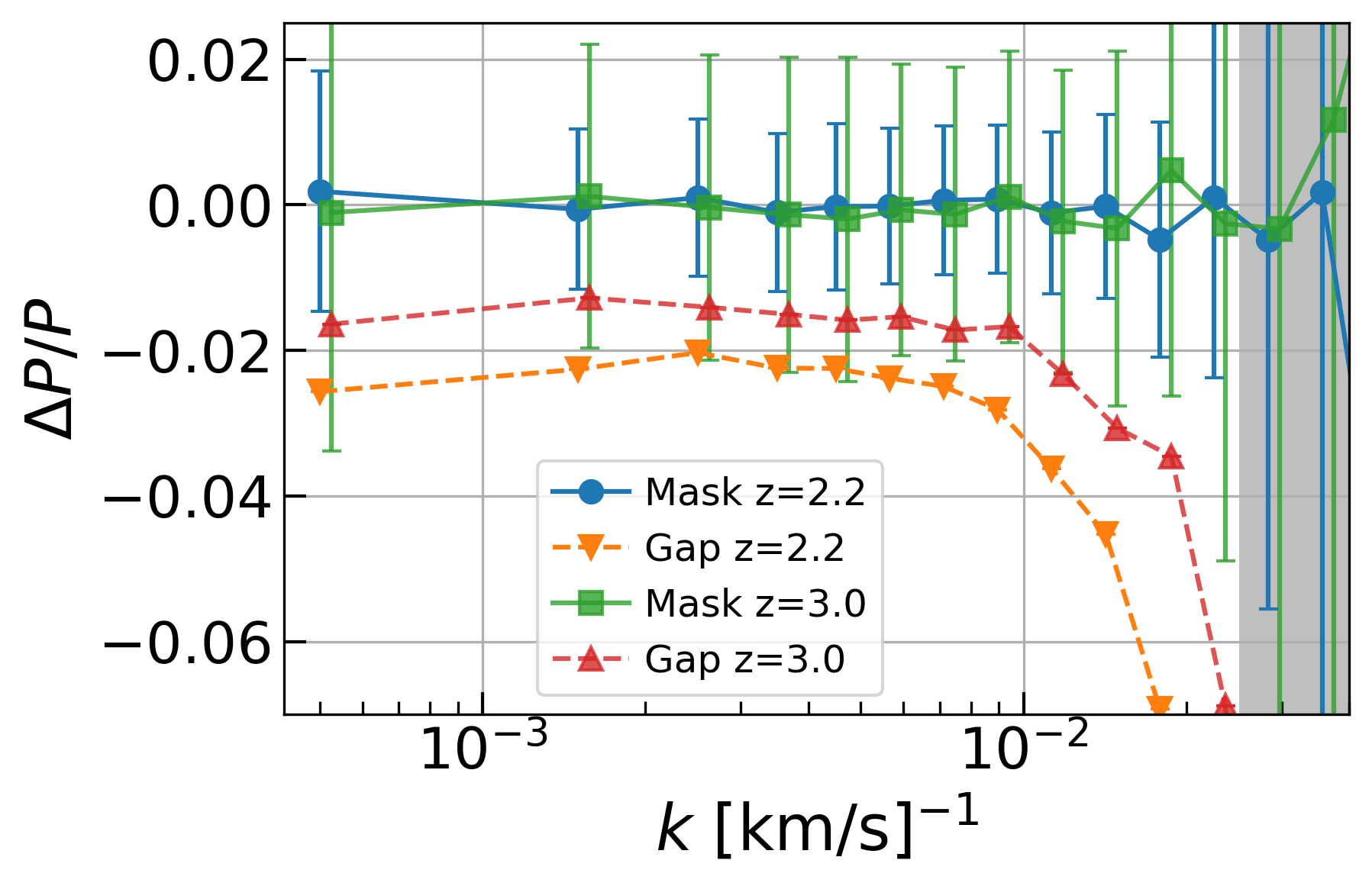}
    \caption{The effect of gaps in $P_{\mathrm{1D}}$ for two redshift bins $z=2.2$ and $z=3.0$. Points labelled "Mask" are corrected measurements by large pixel errors, whereas "Gap" points have unchanged errors. For clarity, we slightly shift $z=3.0$ points and omit error bars from uncorrected results. The masking error propagates to all scales.}
    \label{fig:masking}
\end{figure}

\subsection{Gaps in Spectra}
An advantage of our estimator is that it works in pixel space and can therefore robustly handle missing data in the spectrum. In order to test how masking affects our results, we remove continuous regions from some spectra. We assign 15\% probability of having a high-column density absorber (HCD) in a spectrum, matching  \citet{noterdaeme_column_2012}. If a spectrum has an HCD, we randomly pick a central wavelength and mask 12.5 A$^\circ$ on each side by setting the flux to the mean value ($\delta_F=0$). We apply this random masking procedure to 100,000 spectra. We perform a run where masked pixels are removed by assigning large errors, and another run without removing these pixels for comparison.

Figure~\ref{fig:masking} shows our results in two redshift bins. The masking suppresses power at small scales, but propagates to all scales. The run without any correction yields extremely poor results with $\chi^2 \sim 4,000$. When we correct for these pixels by assigning large errors, the power spectrum estimates yield $\chi^2=234$, close to the original case. This confirms our method is robust against gaps in spectra.

\subsection{Continuum Marginalization}
In a typical pipeline, the real observable flux $f(\lambda)$ is divided by the quasar continuum $C(\lambda)$ to obtain the normalized flux $F(\lambda)$ for each spectrum. Furthermore, flux fluctuations are obtained by diving this with the mean normalized flux $\bar F(\lambda)$. The errors in this process propagate to mostly large scales and are called the continuum errors in Ly-$\alpha$ nomenclature. 

Our quadratic estimator is armed with marginalization capability to suppress these offsets \citep[Appendix B]{slosar_measurement_2013}. By modifying the covariance matrix to $\mathbf{C}' = \mathbf{C} + N \bm t \bm t^{\mathrm{T}}$, where $N$ is large and $\bm t$ is the mode we want to marginalize out, one can show that $\mathbf{C}'^{-1}(\bm{\delta}_F' + \alpha \bm t) \approx \mathbf{C}^{-1} \bm{\delta}_F'$, where the new data vector $\bm{\delta}_F'$ is orthogonal to $\bm t$. This effectively removes any information from data that is in mode $\bm t$.

We will focus our attention to the continuum fitting \citet{slosar_measurement_2013} uses. They fit the flux of each quasar $q$ with $f_q(\lambda_o) = A_q(\lambda_o) C(\lambda_r) \bar F(\lambda_o) (1+\delta_{F, q})$, where $\lambda_o$ is the observed wavelength and $\lambda_r$ is the rest frame wavelength. $C(\lambda_r)$ and $\bar F(\lambda_o)$ are determined globally, whereas the function $A_q$ has two free parameters for each quasar:
\begin{equation}
    A_q(\lambda) = a_q + b_q \frac{\ln \lambda - \ln\lambda_1}{\ln\lambda_2 - \ln\lambda_1},
\end{equation}
where $\lambda_{1,2}$ are the beginning and the end of the forest in a given spectrum. These two quasar specific parameters source most of the error assuming the global functions are more robust.

\begin{figure}
    \centering
    \includegraphics[width=\linewidth]{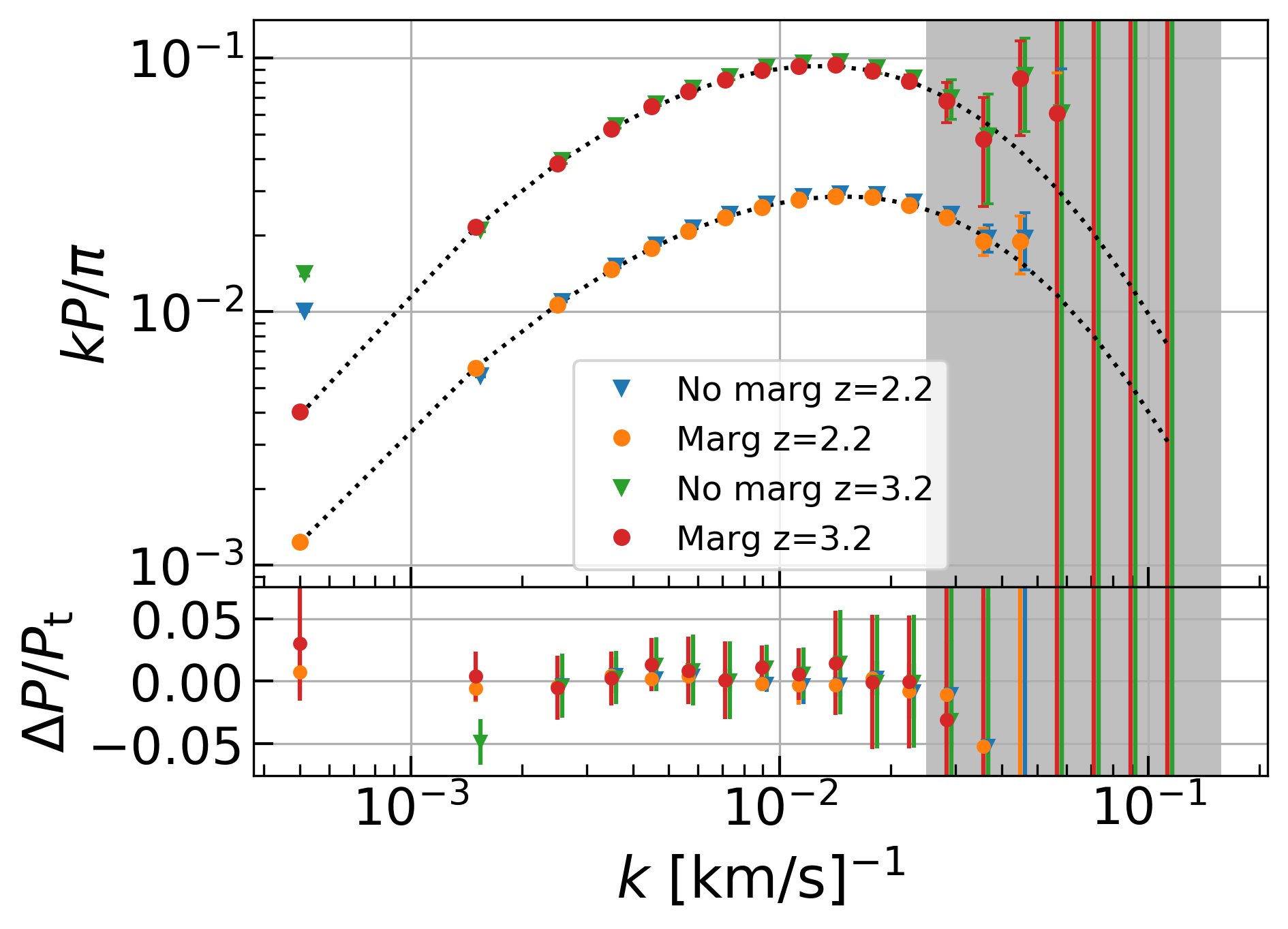} 
    \caption{Average power spectra of eight runs at $z=2.2$ and $z=3.2$ with and without marginalizing continuum errors. The continuum errors highly contaminate the first two $k$ bins at $k=5\times 10^{-4}$ s km$^{-1}$ and $k=1.5\times 10^{-3}$ s km$^{-1}$. The remaining scales are less affected. By marginalizing two continuum modes, we are able to correctly estimate the power at these scales. Average errors of eight runs at $k=5\times 10^{-4}$ s km$^{-1}$ bin increase by 25--45\%. The change in the second bin is insignificant.}
    \label{fig:continuum_power}
\end{figure}

To replicate these continuum errors, we add an error $\eta(\lambda)$ to each quasar's flux fluctuations $\delta_F(\lambda)$. We limit the form of $\eta(\lambda)$ to the equation above in order to perform a controlled test. We randomly generate the two parameters from a Gaussian distribution with $\sigma=0.1$. We run eight independent sets of 100,000 spectra.

We add a large constant $N_0$ to all elements of the covariance matrix to marginalize the amplitude, and $N_1 \bm t \bm t^{\mathrm{T}}$ to marginalize the slope, where $t=\ln(\lambda/\lambda_{\mathrm{Ly}\alpha})$. 

Figure~\ref{fig:continuum_power} presents the average of eight runs. The continuum errors mostly contaminates large scales. The largest scale $k=5\times 10^{-4}$ s km$^{-1}$ bins show big offsets, whereas the following $k$ bin is less affected. When we marginalize these two continuum terms with $N_0=N_1=1000$, results go back to the expected values. As a side effect, marginalization increases the error in the first bin by 25--45\%.

\subsection{Choice of Fiducial}
We already established in the previous sections that using the true power as fiducial yields correct results. In this section, we investigate how different fiducial power influences results by considering two additional cases. First, we run the estimator without any fiducial power.  Second, imagining a realistic pipeline, we fit equation~\ref{eq:pd13_fitting_fn} to these no fiducial results, and employ this best-fitted function as fiducial. In order to run multiple independent realizations, we use eight sub-samples of 100,000 spectra.

\begin{figure}
    \centering
    \includegraphics[width=\linewidth]{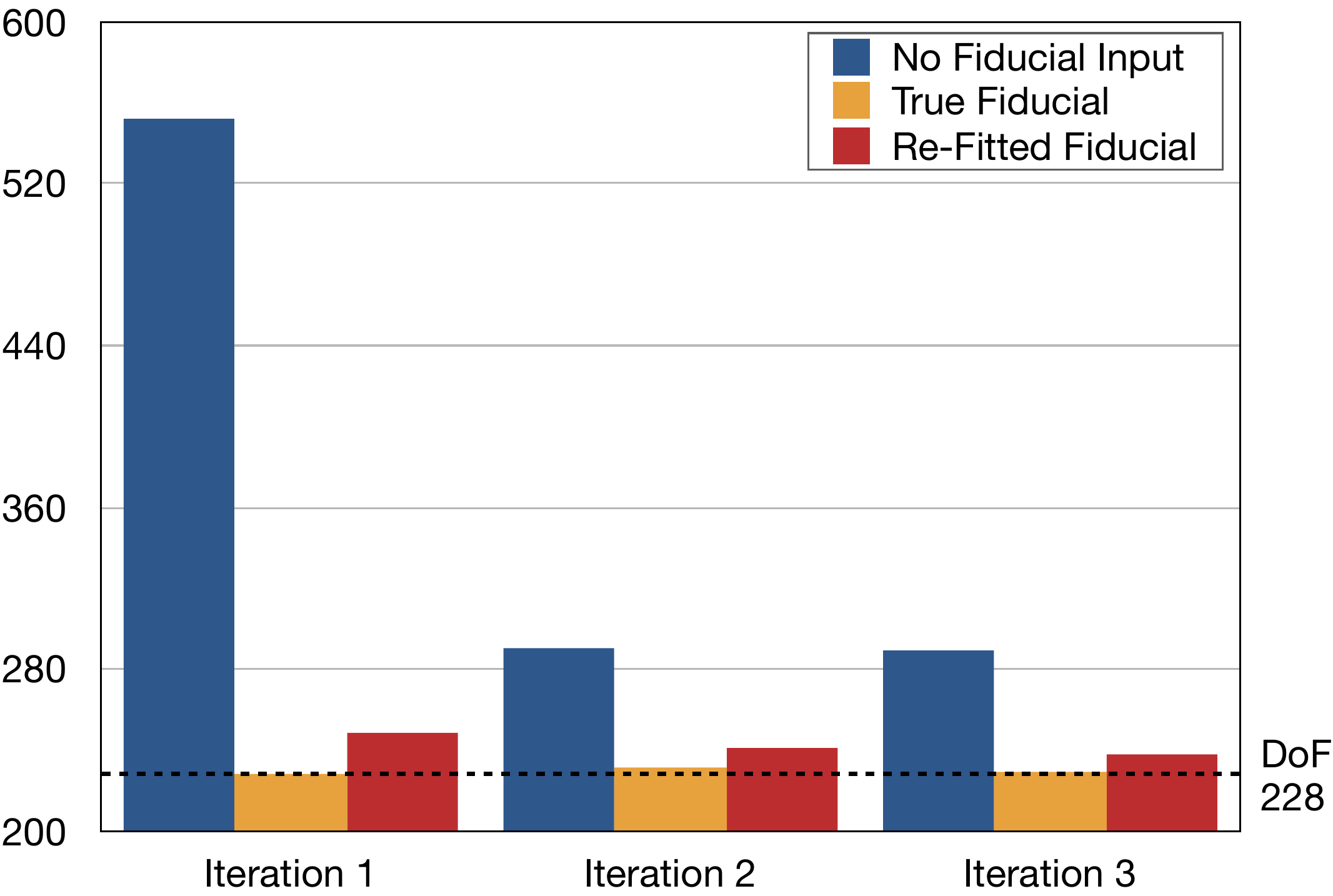}
    \caption{Comparing average $\chi^2$ of eight runs for three different fiducials using respective Fisher matrices. True fiducial yields the correct answer in one iteration and outperforms other cases as expected. Not using any prior naturally starts away from the truth, but converges to the correct power immediately. Using the results from no prior to construct a better fiducial (re-fitted fiducial) decisively surpasses no fiducial case. Repeating this analysis with true Fisher matrix yields the same conclusion. In that case, $\chi^2$ stays constant over three iterations with values close iteration two and three of this figure. This implies that the first iteration yields good $P_{\mathrm{1D}}$ and subsequent iterations correct the error estimates.}
    \label{fig:compare_chisq_ave}
\end{figure}

We compare $\chi^2$ for every case using respective Fisher matrices, which can be seen in figure~\ref{fig:compare_chisq_ave}. True fiducial yields the correct answer in one iteration and is the best estimate as expected. Not using any prior naturally starts away from the truth, but converges to the correct power immediately while being relatively a poor fit. Using the results from no prior to construct a better fiducial decisively outperforms no fiducial case, which we found to hold for all eight runs.

The $\chi^2$ analysis above weights absolute errors with respective Fisher matrices. The differences in these matrices partially source the $\chi^2$ trend in figure~\ref{fig:compare_chisq_ave}. In order to decouple error accuracy from $P_{\mathrm{1D}}$ accuracy, we also calculate all $\chi^2$s using the true Fisher matrix. We find this quantity still distinguishes different fiducials as before with values close to the last two iterations of the initial analysis, and stays approximately flat over three iterations. This indicates that QMLE yields good $P_{\mathrm{1D}}$ at the first iteration even for no fiducial case, but it does not improve at subsequent iterations. Evidently, these iterations mostly correct the error estimates. This behaviour is reasonable given the weights in equation~\ref{eq:cont_limit} stay approximately constant across bins, and therefore are unaffected by band power corrections. The only way to overcome the band power discretization is through integrating a fiducial power.

To summarize, the key points are as follows: 1) even with no fiducial, the power spectrum is correctly recovered, 2) a simple iteration quickly gets the correct Fisher matrix, 3) a good fiducial yields better power spectrum and error estimates.

\subsection{Fisher Matrix Approximation}
\label{sec:fisher_optm}
Each quasar's Fisher matrix requires $N_B(N_B+1)/2$ matrix multiplications, where $N_B$ is the number of total bins to which this spectrum contributes. However, figure~\ref{fig:fisher_zbyz} tells us that only few elements are dominant in the Fisher matrix. Moreover, we see this structure in all cases, i.e. different fiducial powers, masking and continuum marginalization. 

While the sparsity structure of the Fisher matrix appears to be quite generic, the exact details (as to the number of non-zero elements) do appear to depend on the total signal-to-noise of the sample. For the DESI-like data considered in this work, we find that including a single off-diagonal element in the $k-$ and $z-$ directions works very well. Specifically, for every $(k_n, z_m)$ pair we calculate $(k_{n+1}, z_m)$ and $(k_n, z_{m+1})$ elements besides the diagonal. That reduces the number of matrix multiplications to $3N_B$, which will boost the speed. Note that this should also be taken into account in load balancing.

We test this optimisation scheme on 800,000 spectra. We find that the resulting $\chi^2$ is not changed while the average time per iteration decreases by 64\% and the time spent in Fisher matrix calculation per iteration drops by 95\%. 

Even though we do not have an exact prescription for computing the number of non-zero elements, the above discussion suggests a procedure by which one starts to fill in the Fisher matrix from the diagonals outwards and stops when the overall relative contribution is below a chosen threshold.

\begin{table}
    \centering
    \begin{tabular}{r|r|r|r|r}
        $z$ & $P_{\mathrm{1D}}(k_f)$  & Precision & Precision & Precision\\
         & [km s$^{-1}$] & S/N=1 \AA$^{-1}$ & S/N=2 \AA$^{-1}$ & S/N=4 \AA$^{-1}$\\
        \hline
        2.0 & 10.2 & 1.31\% & 0.44\% & 0.22\% \\
        2.2 & 13.4 & 1.08\% & 0.37\% & 0.20\% \\
        2.4 & 17.5 & 1.12\% & 0.40\% & 0.23\% \\
        2.6 & 22.8 & 1.24\% & 0.46\% & 0.27\% \\
        2.8 & 29.5 & 1.46\% & 0.56\% & 0.34\% \\
        3.0 & 37.8 & 1.87\% & 0.74\% & 0.45\% \\
        3.2 & 48.3 & 2.50\% & 0.99\% & 0.62\% \\
        3.4 & 61.4 & 3.34\% & 1.32\% & 0.82\% \\
        3.6 & 77.6 & 4.82\% & 1.89\% & 1.15\% \\
        3.8 & 97.7 & 8.67\% & 3.32\% & 1.98\% \\
        4.0 & 123.0 & 22.00\% & 8.14\% & 4.63\% \\
        4.2 & 153.0 & 95.30\% & 33.30\% & 18.10\% \\
    \end{tabular}
    \caption{Our model power spectrum and precision forecasts for 5-year DESI survey for fixed $k_f=2.5\times 10^{-3}$ s km$^{-1}$. The precision is defined as error divided by signal: $e/P_{\mathrm{1D}}$. For reference, metal contamination constitutes 5--15\% of the Ly-$\alpha$ power in reality.}
    \label{tab:5y_forecasts}
\end{table}

\subsection{5-Year Forecasts}
\label{subsec:forecasts}
We would like to make simple predictions for future DESI performance using our simple spectra. We run two additional cases where S/N=1 \AA$^{-1}$ and S/N=4 \AA$^{-1}$ on 800,000 spectra. In order to speed up our analysis we turn on the Fisher optimisation scheme. We perform only one iteration with true power as fiducial. 

We compare $P_{\mathrm{1D}}$ estimates for different redshifts at a fixed wavenumber $k_f=2.5\times 10^{-3}$ s km$^{-1}$. Our results are summarized in table~\ref{tab:5y_forecasts}. This prediction is under ideal circumstances, hence further complications should be added for accuracy. For example, metal contamination constitutes 5--15\% of the Ly-$\alpha$ forest power \citep{day_power_2019}.

We provide the model power spectrum and the full Fisher matrices in the electronic form. A summary of the Fisher matrices is in table~\ref{tab:fisher_matrices}.

In the noise dominated limit, the covariance matrix is $\mathbf{C} \approx \mathbf{N}$, and so the Fisher matrix scales as $\mathbf{N}^{-2}$. Assuming the noise is uncorrelated and Gaussian with standard deviation $\sigma$, this means that the Fisher matrix is proportional to $\mathbf{F} \propto \sigma^{-4} = \mathrm{(S/N)}^4$ (or the precision scales as (S/N)$^2$), where we substituted the definition S/N$\equiv 1/\sigma$. We confirmed that this scaling holds true for high $k$ values in the Fisher matrix.

\begin{table*}
    \centering
    \begin{tabular}{c|c|c|c|c}
($z_i$, $k_i$)  &  ($z_j$, $k_j$) & $F_1$ & $F_2$ & $F_4$ \\
         (, [s km$^{-1}$]) & (, [s km$^{-1}$]) & [s$^2$ km$^{-2}$] & [s$^2$ km$^{-2}$] & [s$^2$ km$^{-2}$] \\
\hline
(2.0, 5.00$\times 10^{-4}$) & (2.0, 5.00$\times 10^{-4}$) & 6.02867$\times 10^{1}$ & 6.51197$\times 10^{2}$ & 3.66215$\times 10^{3}$ \\
(2.0, 5.00$\times 10^{-4}$) & (2.0, 1.50$\times 10^{-3}$) & 6.63624$\times 10^{0}$ & 6.52238$\times 10^{1}$ & 3.02865$\times 10^{2}$ \\
(2.0, 5.00$\times 10^{-4}$) & (2.2, 5.00$\times 10^{-4}$) & 2.12509$\times 10^{1}$ & 2.30833$\times 10^{2}$ & 1.33759$\times 10^{3}$ \\
(2.0, 1.50$\times 10^{-3}$) & (2.0, 1.50$\times 10^{-3}$) & 6.54364$\times 10^{1}$ & 6.00179$\times 10^{2}$ & 2.47973$\times 10^{3}$ \\
(2.0, 1.50$\times 10^{-3}$) & (2.0, 2.50$\times 10^{-3}$) & 5.99838$\times 10^{0}$ & 5.43360$\times 10^{1}$ & 2.19731$\times 10^{2}$ \\
(2.0, 1.50$\times 10^{-3}$) & (2.2, 1.50$\times 10^{-3}$) & 1.97400$\times 10^{1}$ & 1.75264$\times 10^{2}$ & 6.90261$\times 10^{2}$ \\
(2.0, 2.50$\times 10^{-3}$) & (2.0, 2.50$\times 10^{-3}$) & 6.38029$\times 10^{1}$ & 5.69068$\times 10^{2}$ & 2.25086$\times 10^{3}$ \\
(2.0, 2.50$\times 10^{-3}$) & (2.0, 3.50$\times 10^{-3}$) & 5.83666$\times 10^{0}$ & 5.28085$\times 10^{1}$ & 2.13079$\times 10^{2}$ \\
(2.0, 2.50$\times 10^{-3}$) & (2.2, 2.50$\times 10^{-3}$) & 1.92028$\times 10^{1}$ & 1.65626$\times 10^{2}$ & 6.24662$\times 10^{2}$ \\
(2.0, 3.50$\times 10^{-3}$) & (2.0, 3.50$\times 10^{-3}$) & 6.29898$\times 10^{1}$ & 5.69389$\times 10^{2}$ & 2.29742$\times 10^{3}$ \\
(2.0, 3.50$\times 10^{-3}$) & (2.0, 4.50$\times 10^{-3}$) & 5.74635$\times 10^{0}$ & 5.31461$\times 10^{1}$ & 2.21748$\times 10^{2}$ \\
(2.0, 3.50$\times 10^{-3}$) & (2.2, 3.50$\times 10^{-3}$) & 1.89681$\times 10^{1}$ & 1.66002$\times 10^{2}$ & 6.39374$\times 10^{2}$ \\
(2.0, 4.50$\times 10^{-3}$) & (2.0, 4.50$\times 10^{-3}$) & 6.21816$\times 10^{1}$ & 5.77621$\times 10^{2}$ & 2.43005$\times 10^{3}$ \\
(2.0, 4.50$\times 10^{-3}$) & (2.0, 5.65$\times 10^{-3}$) & 5.85085$\times 10^{0}$ & 5.58077$\times 10^{1}$ & 2.44502$\times 10^{2}$ \\
(2.0, 4.50$\times 10^{-3}$) & (2.2, 4.50$\times 10^{-3}$) & 1.87450$\times 10^{1}$ & 1.68881$\times 10^{2}$ & 6.78999$\times 10^{2}$ \\
(2.0, 5.65$\times 10^{-3}$) & (2.0, 5.65$\times 10^{-3}$) & 8.19711$\times 10^{1}$ & 7.90781$\times 10^{2}$ & 3.53432$\times 10^{3}$ \\
(2.0, 5.65$\times 10^{-3}$) & (2.0, 7.11$\times 10^{-3}$) & 6.07740$\times 10^{0}$ & 6.05452$\times 10^{1}$ & 2.85148$\times 10^{2}$ \\
 \vdots & \vdots & \vdots & \vdots & \vdots \\
    \end{tabular}
    \caption{Our Fisher matrix forecasts for 5-year DESI survey for different spectral qualities. The subscript in $F$ refers to the S/N value. The full table and model power spectrum can be found in the electronic submission of this article.}
    \label{tab:fisher_matrices}
\end{table*}

\section{Discussion}
\label{sec:discussion}
We would like to start our discussion by highlighting the differences between QMLE and FFT. As we have shown in section~\ref{subsec:cont_limit}, QMLE finds inverse variance weighted averages across bins. Measuring deviations from a baseline power further lessens the averaging, so that QMLE yields near exact $P_{\mathrm{1D}}(k)$. On the other hand, FFT estimator computes simple averages. For wide enough bins, this does not equal to $P_{\mathrm{1D}}(k)$. Therefore, it is important to note that these two methods will not fully agree unless special circumstances are met, because they compute mathematically different quantities. 

Systematic errors are the significant uncertainty source in $P_{\mathrm{1D}}$ analysis. We have considered three such error sources: 1) high-column density absorbers, 2) metal contamination and 3) continuum fitting. Using our method we showed that HCDs can be masked without further complications. Modelling their contribution can now be solely a theoretical question as they are biased tracers in reality. However, metal contamination is more complicated as they are not easily separable from data like HCDs. Metal contamination estimates will inevitably bring statistical errors; and these errors can be fairly higher than our simple forecasts. Nevertheless, all methods face this challenge. Third, continuum errors can be more complicated than our simple model, but we believe these errors can still be marginalized with careful investigation. Even though marginalization washes out some information, it still enables us to keep the largest scales in $P_{\mathrm{1D}}$. Furthermore, DESI will supply plenty of spectra to study and better understand these errors in the future. 

Two additional error sources not considered here are uncertainties in the estimates of the resolution
and the noise. Although we used a simple Gaussian form of the spectrograph resolution and 
a diagonal noise matrix, the QMLE formalism allows us trivially extend this to more complicated forms, such as 
described in \cite{bolton_spectro-perfectionism_2010}. We defer a complete analysis, including a sensitivity
analysis to resolution and noise misestimations to future work.

A disadvantage of our method over a direct FFT method is the additional computational time. However, we demonstrated that these calculations are now practical even for surveys of the scope of DESI, as we were able to perform multiple runs of full DESI-like data in a small cluster of 30 nodes with 24 cores each. We also introduced an optimisation scheme that brings down the CPU time significantly. We think this cost is reasonable given QMLE's capacity to overcome Ly-$\alpha$ forest specific challenges. Further computational speedups such as using GPUs are likely possible; we defer these to later work.

\section{Summary}
\label{sec:summary}
The Ly-$\alpha$ forest has emerged as a unique and competitive tool to investigate the large-scale structure of the universe. This technique can probe cosmological parameters on large scales, while being sensitive to the thermal state of the IGM, neutrino masses and new dark matter models on small scales. This small scale physics enriches the 1D and 3D power spectrum of the Ly-$\alpha$ forest.

In this work, we studied the optimal quadratic estimator (QMLE) for $P_{\mathrm{1D}}$ by first formulating its generic and Ly-$\alpha$ specific expressions, offered an analytic continuum limit formula to depict what QMLE calculates and a straightforward way to bootstrap its results. Under the simplest terms, QMLE finds an inverse variance weighted average power for each bin. We then outlined the implementation in detail by providing our step-by-step algorithm. We here underline two of our decisions that mitigates the systematics and numerical instabilities: 1) we fed smoothed estimates to the covariance matrix, and 2) we picked a convergence criteria that weights changes between iterations by the error estimates. Then, we described our synthetic spectra and provided analytic expressions for what they construct. These analytic expressions were crucial to ascertain the performance of QMLE.

We generated DESI-oriented synthetic spectra in order to perform comprehensive tests. These mocks assumed constant resolution ($R=3200$) and noise at all wavelengths, and limited the quasar redshift range to $z_{\mathrm{qso}} \in [2.1, 4.4]$ with pixel width $c \Delta\ln\lambda = 30$ km s$^{-1}$. A summary of our findings is as follows:
\begin{enumerate}
    \item Using 800,000 these spectra with S/N=2 \AA$^{-1}$, we first showed that the power spectrum could be accurately measured with 1\% precision. This number declined towards higher redshifts due to diminishing statistics. This proved the absence of biases in our algorithm.
    
    \item We randomly masked $\Delta\lambda=25$ \AA\, region of 15\% of spectra, and showed that this masking badly affected the measurement when untreated, and proved QMLE was robust against its effects on 100,000 mocks. As real analyses would mask bad pixels and high density absorbers, this strength is invaluable to $P_{\mathrm{1D}}$ analysis.
    
    \item We introduced continuum errors by adding wavelength-dependent error $\eta(\lambda) \sim 10\%$ to flux fluctuations $\delta_F$. This error function $\eta(\lambda)$ had two independent parameters for each quasar: amplitude and slope. Using eight independent runs with 100,000 spectra each, we demonstrated that these continuum errors contaminate first two $k$ bins, but QMLE could marginalize out the noisiest modes, and recover these scales. 

    \item We presented, both analytically and by using 100,000 spectra, how a baseline model improved the accuracy. This fiducial power enables us to integrate over bins and to lessen the discretization of band powers. Our proposed unbiased procedure to find this fiducial power is to first measure the power spectrum without any input, then find the best-fitted analytic function on these results. We showed this feature significantly boosted our measurements.
    
    \item Finally, we found that computation was mostly spent on Fisher matrix calculation. We also found that Fisher matrix had a simple structure, which allowed us to come up with an optimisation scheme that reduced the computation time by 60\%.
\end{enumerate}

This work represents an initial step towards analyzing the upcoming Lyman-$\alpha$ datasets with surveys like DESI. The quadratic estimator formalism allows us to optimally use all the available data (even with varying S/N) and to effectively mitigate systematic errors like those arising from an imperfect continuum estimate. We address several practical issues with implementing a QMLE power spectrum code for the Lyman-$\alpha$ forest. Future work will use these results and the codes described here to analyze both existing high-resolution data as well as upcoming DESI data.

\section{Data availability}
The data underlying this article are available in the article and in its online supplementary material.

\section*{Acknowledgements}
We thank Michael Walther, Nathalie Palanque-Delabrouille, Christophe Y\` eche, Vid Ir\v si\v c and  DESI Lyman-$\alpha$ working group for useful discussions. We also thank the referee for their detailed comments and suggestions.

NGK and NP are supported by DOE, reference number DE-SC0017660. AFR acknowledges support by an STFC Ernest Rutherford Fellowship, grant reference ST/N003853/1, and by FSE funds trough the program Ramon y Cajal (RYC-2018-025210) of the Spanish Ministry of Science and Innovation.

This is a pre-copyedited, author-produced PDF of an article accepted for publication in \textit{Monthly Notices of the Royal Astronomical Society} following peer review. The version of record \textit{Naim G\" oksel Kara\c cayl\i, Andreu Font-Ribera, Nikhil Padmanabhan, Optimal 1D Ly-$\alpha$ Forest Power Spectrum Estimation I: DESI-Lite Spectra, MNRAS, staa2331} is available online at: \url{https://doi.org/10.1093/mnras/staa2331}.




\bibliographystyle{mnras}
\bibliography{lya_refs} 



\appendix
\section{Saddle Point Approximation for the Mean Flux of Log-normal Mocks}
\label{app:saddle}
We can find a closed analytic form for the mean flux in equation~\ref{eq:theo_mean_flux} using saddle point approximation assuming the integrand is large around its maximum. We start by finding the derivatives of the function in the exponential:
\begin{align}
     \phi(\delta) &\equiv - \frac{\delta^2}{2\sigma^2} - x e^{2a\delta}, \\
     \phi'(\delta) &= -\frac{\delta}{\sigma^2} - 2axe^{2a\delta}, \\
     \phi''(\delta) &= -\frac{1}{\sigma^2} - 4a^2xe^{2a\delta},
\end{align}
where the $z$ dependence of $x(z)$ and $a(z)$ are suppressed for clarity. Solving for the maximum $\phi'(\delta_*) = 0$ yields $-\delta_* = 2a\sigma^2xe^{2a\delta_*}$, for which the solution is given by the Lambert $\mathcal{W}$ function. 
\begin{equation}
    \delta_*(z) = -\frac{1}{2a}\mathcal{W}\left(4a^2\sigma^2x\right)
\end{equation}
Note that $\delta_*<0$. Substituting this solution back into $\phi$ and $\phi''$ yields:
\begin{align}
    \phi(\delta_*) &= \frac{\delta_*}{2a\sigma^2}(1-a\delta_*), \\
    \phi''(\delta_*) &= - \frac{1-2a\delta_* }{\sigma^2}.
\end{align}
Now we can approximate the integration around the maximum.
\begin{align}
    \bar F(z) &\approx \frac{1}{\sigma\sqrt{2\pi}} \int_{\delta_*-\epsilon}^{\delta_*+\epsilon} \exp{\left[\phi(\delta_*) + \frac{(\delta-\delta_*)^2}{2}\phi''(\delta_*) \right]} d\delta \\
    &\approx \frac{e^{\phi(\delta_*)}}{\sqrt{2\pi\sigma^2}} \int_{-\infty}^{\infty} \exp{\left[\phi''(\delta_*) \frac{\delta^2}{2} \right]} d\delta = \frac{e^{\phi(\delta_*)}}{\sqrt{- \phi''(\delta_*) \sigma^2}},
\end{align}
and note that $-\phi''(\delta_*) \sigma^2 = 1-2a(z)\delta_*(z)$. We can further define $d_*(z)=a(z)\delta_*(z)$ and $\sigma(z)=a(z)\sigma$ to simplify this expression.
\begin{equation}
    \bar F(z) \approx \left[1-2d_*(z)\right]^{-1/2} \exp\left\{\frac{d_*(z)[1-d_*(z)]}{2\sigma^2(z)}\right\}
\end{equation}
As noted in the main section, this approximation starts to deviate from the truth at $z\lesssim 3$ up to 4\%. It would be interesting to use this function as a fitting template for the real data.


\bsp	
\label{lastpage}
\end{document}